**Statistical Study on the Sources of Jovian Decametric Radio Emissions Based on the Radio Observations of Remote Instruments**

Ruobing Zheng[1,2], Yuming Wang[1,2,3,*], Xiaolei Li[1,3], Chuanbing Wang[1,2], and Xianzhe Jia[4]

[1]Chinese Academy of Sciences Key Laboratory of Geospace Environment, School of Earth and Space Sciences, University of Science and Technology of China, Hefei 230026, China;

[2]Chinese Academy of Sciences Center for Excellence in Comparative Planetology, University of Science and Technology of China, Hefei 230026, China;

[3]Mengcheng National Geophysical Observatory, School of Earth and Space Sciences, University of Science and Technology of China, Hefei 230026, China;

[4]Department of Climate and Space Sciences and Engineering, University of Michigan, Ann Arbor, MI 48109-2143, USA

[*]Corresponding author (ymwang@ustc.edu.cn)

**Abstract.** To better understand the physical processes associated with Jovian decametric (DAM) radio emissions, we present the statistical study of DAMs and inferred characteristics of DAM sources based on multi-view observation from Wind and STEREO spacecraft. The distribution of the apparent rotation speed of DAMs derived from multiple spacecraft suggests that the rotation speed of Io-related DAMs is in range of $0.15 - 0.6\ \Omega_j$ and that of non-Io-DAMs is between $0.7 - 1.2\ \Omega_j$. Based on the method of Wang et al. (2020), we locate the sources of the DAMs and infer their emission angles and associated electron energies. The statistical results show that the DAM source locations have three preferred regions, two in the southern hemisphere and one in the northern hemisphere, which is probably caused by the non-symmetrical topology of Jupiter's magnetic field. The difference between Io-DAM source footprints and Io auroral spots changes with the Io's position in longitude, consistent with the previous results from Hess et al. (2010), Bonfond et al. (2017) and Hinton et al. (2019). In addition, the emission angles for non-Io-DAMs are smaller than that for Io-DAMs from the same source regions and all the emission angles range from $60^\circ$ to $85^\circ$. Correspondingly, the electron energy is mainly distributed between 0.5 and 20 keV.

# 1. Introduction

Jupiter auroral radio emission has various components spanning the frequency range below 10 kHz to above 3 GHz and is the mainly intensive non-thermal radiation generated in the Jovian inner magnetosphere and upper ionosphere (Zarka, 1998). Firstly discovered by Burke & Franklin (1955) and ranges from 1 MHz to 40 MHz, decametric (DAM) radio emission is the most intensive radio emission from Jupiter. It was shown that the mechanism for generating the Jovian radio emission is the Cyclotron Maser Instability (CMI), a theory widely used in planetary radio emission (e.g., Wu & Lee, 1979; Zarka, 1998; Huff et al., 1988; Kurth et al., 2005; Treumann, 2006), and the loss-cone electron distribution has been recently proven by Juno in-situ observations to be the one driving the CMI (e.g., Louarn et al., 2017, 2018). Based on the loss-cone driven CMI theory, Hess et al. (2008, 2010) established an emission angle function related to active electron energy and Jovian magnetic field. This function has become the core of several methods, e.g., the ExPRES (Exoplanetary and Planetary Radio Emission Simulator, see Louis et al., 2019) and the method developed by Wang et al. (2020, 2022), to infer the sources and characteristics of DAM emissions.

Abundant observations and CMI theory suggested that DAM radiates along a hollow emission cone with a narrow cone wall. The emission half angle is between 60°–90° (Queinnec & Zarka, 1998, 1998S; Hess et al., 2008, 2014) and the thickness of cone wall is about 1°–2° (Kaiser et al., 2000; Panchenko & Rucker, 2016). Based on the CMI theory and the refraction effect in the source region, the emission angle varies with the radiation frequency, and the DAM appears like arc-shape in the time-frequency spectrum lasting from minutes to hours (Carr et al., 1983; Zarka, 1998; Hess et al., 2008). In the observer's view, the dawn-side emission is vertex-early (VE) arc and the dusk-side emission is vertex-late (VL) due to the geometry of the emission cone angle. The emission in the northern hemisphere is usually right-polarized, while the emission in the southern hemisphere is usually left-polarized due to the angle between the radiation wave vector and the Jupiter magnetic field vector. Based on the above characteristics of the DAM arc shape in the spectrum and the polarization of wave, DAM emissions can be classified into four regions A (the northeastern quadrant), B (the northwestern quadrant), C (the southeastern quadrant), and D (the southwestern quadrant) relative to an observer (see Fig. 2 in Ray & Hess, 2008), and a statistical study based on ground-based radio observations from Nançay Decameter Array (Marques et al., 2017) has counted the distribution of each DAM type in the central meridian longitude (CML) - Io phase diagram.

In previous studies on the DAM source location, remote sensing observations combined with modeling have been commonly used before the advent of in-situ observations. Imai (2002) modeled the high-frequency part of the Io-DAM spectrum based on the assumption that the emission cone is a linear and thin structure along Jupiter's tilted dipole magnetic field and rotates synchronously with Io, then the emission angle and the magnetic field lines associated with the source region can be calculated by model. Hess et al. (2010) fitted the Io-DAM arcs in the radio spectrum based on the emission angle function and summarized an empirical relation between the resonant electron energy and Io in System III longitude. Martos et al. (2020) studied Io-DAM events from four regions observed by Juno/WAVEs. They estimated the source location, the emission angles and corresponding electron energy through for the four events. Imai et al. (2008, 2011) used Cassini/RPWS to analyze the occurrence probability of each frequency of the non-Io-DAMs with CML and found that the non-Io-A and non-Io-B source locations are projected at 180° $\pm$ 10° of System III longitude. More recently, Imai et al. (2017) calculated the probability distribution of DAMs with the latitude range -21° to 15° of the Juno polar orbit, and the results showed two peaks of the probability of non-Io-B occurring in the interval of 170° to 140° and 110°.

Recently, Wang et al. (2020) developed a new method to infer DAM source locations and emission characteristics based on remote radio dynamic spectra. Wang's method can just input the observed DAM arcs in a range of frequency and derive all the associated source field lines, which is different from other methods. Here, we investigate 111 DAM events observed by multiple spacecraft, i.e., Wind, STEREO-A and STEREO-B , in the 1-AU orbit, and apply the method of Wang et al. (2020) to study their statistical properties. In Section 2, we introduce the selection and classification of DAM events, and present the observational statistical results of the events. In Section 3, we briefly describe the Wang's method, and use some examples in different types of DAMs to illustrate the method. In Section 4, we show the statistical results of the inferred source locations, emission angles and associated electron energies of the DAM events, and then, in Section 5 we give the summary and discussion.

# 2. Observations and DAM Events

The combination of Wind and STEREO missions can provide stereoscopic observation of Jovian radio emission. The Wind spacecraft was launched in November 1994 and is currently placed in a halo orbit around the L1 Lagrange point. The Wind carries WAVES instrument (J.-L. Bougeret et al., 1995) for plasma and radio wave in space, in which radio receiver band 2 (RAD2) ranges from 1.075 MHz to 13.825 MHz with 50 kHz frequency step. The STEREO consists of twin spacecraft STEREO-A (ST-A) and STEREO-B (ST-B), launched in October 2006, which orbited the Sun and drifted respectively from the Earth in opposite directions. The STEREO/WAVES instrument (J. L. Bougeret et al., 2008) provides the high frequency dual sweeping receiver (HFR) operated in the 0.125-16.075 MHz range. All these WAVES instruments cover the main frequency of DAM emissions though the higher part of emissions in frequency is missed.

## 2.1 Selection of Events

Thanks to the stereoscopic observations from multiple spacecraft, DAMs can be easily distinguished from solar radio burst due to the large time difference among the time-frequency spectra from the different spacecraft. The configuration of the Wind, ST-A and ST-B could be up to about 21 degrees with respect to Jupiter. Thus, the time difference of observed DAM could be up to about few hours mainly due to the rotation of the radio source. Reversely, with the known angular separation of the observers with respect to Jupiter, we can estimate the rotational speed of the radio source of a DAM by measuring the time difference between the observers.

We manually check the radio dynamic spectra of the three WAVES instruments from 2008 to 2014 September, when ST-B lost communication, to select DAM events. The selection criteria are as follows. (1) A selected DAM should be observed asynchronously by at least two spacecraft and should have similar morphology in a broad frequency range. Considering the limit of the thickness of the emission wall (Panchenko & Rucker, 2016), the separation angle of the two observing spacecraft should be greater than 2.0° with respect to Jupiter. (2) The radio intensity observed by Wind is generally larger than 0.2 dB (according to our processing procedure, the threshold may vary among different procedures), and that observed by STEREO is typically larger than 2 dB, which means we select intense DAM events. (3) The events are isolated and do not contain complex structures in the spectrum. Figure 1 shows two examples for exclusion. Both of them consist of a set of arcs generated by multiple reflections of Alfvén waves. The groups of arcs are difficult to correspond to one another in multiple observed spectrums. Finally, we selected 111 DAM events with a bias towards strong, long-lasting and isolated DAMs (see Table 1), among which 85 events were observed by all of the three spacecraft. For the other 26 events observed by only two spacecraft, the missing spacecraft is not between the other two spacecraft with respect to Jupiter to mostly guarantee the continuous emission of the DAM.

## 2.2 Classification of DAM Types

To distinguish Io-DAMs and non-Io DAMs, the apparent rotation speed, $\Omega_{DAM}$, of a DAM emission is used. It can be calculated by the angular separation, $\theta$, between two spacecraft, the time difference of light travel, $dt_l$, from Jupiter to the two spacecraft and the time difference of the observed DAM features, $dt = t_2 - t_1$, in which $t_1$ and $t_2$ are the times of the DAM in the radio dynamic spectra of the two spacecraft, respectively, i.e.,

$$\Omega_{DAM} = \frac{\theta}{dt - dt_l} \quad (1)$$

The main error source is the measure time $t_1$ and $t_2$. Since the time resolution of the used WAVES spectra is one minute, the uncertainties in $t_1$ and $t_2$ are set to be $\Delta t_1 = \Delta t_2 = 2$ min. Thus, the uncertainty of the calculated $\Omega_{DAM}$ can be estimated by the absolute error transfer formula as follows

$$\Delta\Omega_{DAM} = \left|\frac{\partial\Omega_{DAM}}{\partial t_1}\right|\Delta t_1 + \left|\frac{\partial\Omega_{DAM}}{\partial t_2}\right|\Delta t_2 = \left|\frac{\theta}{(dt - dt_l)^2}\right|(\Delta t_1 + \Delta t_2) \quad (2)$$

For a DAM observed by all of the three spacecraft, we use the two spacecraft with the largest separation angle to derive $\Omega_{DAM}$.

The DAM event observed on 2010 July 10 in Figure 2 is an example. There are three vertex-early arcs in the spectra from Wind/WAVES and ST-A and B/WAVES. The time corresponding to the apex of arc is marked in red, where the time delay, excluding light travel time, between Wind and ST-A is the longest, about 85.6 min, and the angle between the two spacecraft with respect to Jupiter is about 14.44°. According to Equation 1, the apparent rotation speed of DAM is calculated to be approximately $(0.281 \pm 0.013)\ \Omega_J$ while the rotation speed of Io ($\Omega_{Io}$) is about 0.23 $\Omega_J$, in which $\Omega_J$ is the self-rotation speed of Jupiter. It suggests that this DAM emission is Io related event.

Figure 3(a) shows the apparent rotation speeds of the 111 events. The data points mainly distribute between (and roughly around) $\Omega_{Io}$ (blue line) and $\Omega_J$ (gray line). The number of the sample is relatively small from 2011 to 2013 when the solar activity is high because DAM emissions in the radio dynamic spectra were contaminated by solar radio bursts. Figure 3(b) shows the histogram (in bins of 0.05 $\Omega_J$) of the rotation speeds. A double-peak distribution could be found. One peak locates around 0.25-0.3$\Omega_J$ and the other around 0.95-1.0 $\Omega_J$. Based on the distribution, we determine that the 94 DAM events with $0.15\,\Omega_J < \Omega_{DAM} < 0.6\,\Omega_J$ are Io-DAMs (blue bars), and the 15 events with $0.7\,\Omega_J < \Omega_{DAM} < 1.2\,\Omega_J$ non-Io-DAMs (orange bars). The uncertainty of $\Omega_{DAM}$ is considered in the determination of the Io or non-Io DAMs. Two events sitting in the range of 0.6-0.7 $\Omega_J$ are ambiguous ones as indicated by gray bars in the histogram. Based on the rotation speeds, we may conclude that the occurrence probability of isolated strong Io-DAMs is about 6.3 times of that of isolated strong non-Io-DAMs. The rotation speed of Io-DAM is typically larger than that of Io, while non-Io-DAM rotates slower than the Jupiter.

Then we further classify the DAMs into A, B, C, and D types (e.g., Ray & Hess, 2008). A DAM with a vertex-early/late arc in the radio dynamic spectrum is from the western (B, D)/eastern (A, C) hemisphere. However, it's hard to determine a DAM originating from northern (A, B) or southern (C, D) hemisphere without the polarization information. Alternatively, we determine from which hemisphere the emission is more likely to occur based on the characteristic of DAMs in the CML-Io phase plane (Marques et al., 2017). The four types of Io-DAMs occupy different regions in the plane of the observer's CML and the Io phase due to the constraint of Io and active field lines in the source region, though Io-A and Io-C (also Io-B and Io-D) are more or less overlapped in the CML-Io Phase diagram. Similarly, all types of non-Io-DAMs have specific regions distributed on the CML. The difference is that they are uniformly distributed over the Io phase due to being Io-independent. Meanwhile, the tilt of the Jovian magnetic dipole axis is also considered. If the magnetic northern pole is towards the observer during a DAM, we incline to believe that the DAM is from the northern hemisphere.

Based on the CML-Io Phase plane from the ground-based observation, to ensure the classification, we get the CML-Io Phase plane of the 102 events which can be observed by Wind spacecraft near the Earth. The results are shown in Figure 4, where 89 events (with colored markers) can be identified in the range of 10% occurrence probability (see Table 4&5 in Marques et al., 2017) for each type. The CML is mainly distributed in $0^{\circ}-180^{\circ}$ and the Io phase is concentrated in two intervals of $90^{\circ}-120^{\circ}$ and $240^{\circ}-270^{\circ}$, corresponding to the dawn and dusk sides. In the southern hemisphere, there are 15 Io-C events, 60 Io-D events, 10 non-Io-C events and one non-Io-D event. In the northern hemisphere, there are only 3 non-Io-A events. Since the maximum radio frequency of DAM in the northern hemisphere is generally higher than that in the southern hemisphere (e.g., Genova & Aubier, 1985; S. L. G. Hess et al., 2011; Marques et al., 2017) , we can only observe the low frequency parts of northern emission arcs limited by spacecraft instruments, which probably causes exclusions during the initial selection. Thus, the most of the identified events are from the southern hemisphere (i.e., C and D in Fig. 4).

According to the above classification, the sample is in lack of types of A and B, we randomly select 10 events in the northern hemisphere from the database (http://cdsarc.u-strasbg.fr/viz-bin/qcat?J/A+A/604/A17) published by Marques et al. (2017) as a supplement, which were observed by the Nançay Decameter Array (NDA) with frequency range of 10 to 40 MHz . Among the 10 events, 5 ones are Io-A and the others are Io-B. Similar to the previous filter criteria, the added events are strong and isolated arcs in the NDA radio spectrum. The 10 additional events are listed at the bottom of Table 1.

# 3. Tracking Sources of DAMs

## 3.1 Method

The method used to infer the DAM source is from Wang et al. (2020). The main idea is to use the properties of observed DAM arcs to constrain the quadrant of Jupiter and search all the magnetic field lines from the quadrant to find source regions (or the active field lines) that satisfy the time-frequency drift pattern of the observed DAM arcs, which is described by the function (Equation 3) given in Hess et al. (2008),

$$\theta = cos^{-1}\left(\frac{v}{Nc}\sqrt{1-\frac{f}{f_{ce,max}}}\right) \quad (3)$$

where $f$ is observed frequency, approximated to the electron cyclotron frequency, and $f_{ce,max}$ is the maximal electron cyclotron frequency at the active field line footprint (AFT), here the value of $f_{ce,max}$ is set to be the value at 900 km altitude for Io UV footprint (Bonfond et al., 2009), $v$ is resonant electron velocity and $N$ is the refraction index which is approximated to 1 due to strongly magnetized plasma in Jovian magnetosphere, i.e., $f_{pe}/f_{ce} < 0.1 - 0.2$ (e.g., Bagenal, 1994). Eq.3 is based on electron loss-cone driven CMI (e.g., Hess et al. 2008, 2010), which has been proven to be the case in Jovian magnetosphere by Juno in-situ observations (e.g., Louarn et al., 2017, 2018; Louis et al., 2020).

Based on previous studies, the following constraints are applied when we search the active field lines: (1) cone wall thickness is less than 2°; (2) emission angle ranges from 55° to 90°; and (3) electron velocity is greater than 0.05c (or the energy > 0.02 keV). Jupiter's magnetic field is modeled by the internal magnetic field using the 'JMR09' model (Connerney et al., 2018), with an extended magnetic disc taking into account by using the equation of Giampieri & Dougherty (2004). In addition, we use a geodesic polyhedron with the 1/15.4 flattened surface of 1 $R_J$ to trace magnetic field lines (see Fig. 2 in Wang et al., 2020).

In the following sub-sections, we illustrate the method with three examples, which belong to the above identified Io-C, non-Io-A and non-Io-C types, respectively. One of Io-D events was already presented in Wang et al. (2022).

## 3.2 Io-C Event on 2008 January 24

The Io-C event on 2008 January 24 was observed by Wind/WAVES, ST-A/WAVES and ST-B/WAVES (see Figure 5), with observation time from 02:27 to 03:19 UT excluding light travel time. The portion of arc in radio spectra from 5 to 9 MHz is selected for the method, and source location at first observed time are shown in the right panels of Figure 5. The distance of top of field lines (called M-shell value) ranges from 2.5 to 11.7 $R_J$ which crosses the Io's orbit ≈ 5.9 $R_J$. To obtain more intuitive results, the footprints (FPs) of the active field lines on the Jupiter surface are shown in Figure 6. Here we define the lead angle as the longitude difference between the FPs and the footprint of Io (IFP). The averaged longitude of the FPs drift from about 298° to 319° while IFP in southern hemisphere drift from 306° to 328° with the lead angle about 8°~ 9°. The rotation speed of the FPs is about $\Omega_{FP} \approx (0.32 \pm 0.10)\ \Omega_J$ and that of IFP is about $\Omega_{IFP} \approx 0.29\ \Omega_J$ in the inertial coordinates. Compared to the angular rotation speed of DAM in observed spectra which is about $\Omega_{DAM} \approx (0.24 \pm 0.02)\, \Omega_J$, the rotation speed of the DAM source is faster.

Figure 7 illustrates the emission angle and the resonant electron energy during the multi-view observations. The average of emission angle has a small increase from $81^{\circ}$ to $83^{\circ}$. This is probably the reason causing the apparent rotation speed $\Omega_{DAM}$ smaller than that of footprints $\Omega_{FP}$. Accordingly the averaged energy of active electron decreases from 1.50 keV to 1.06 keV.

## 3.3 Non-Io-A Event on 2010 August 6

The non-Io-A event was observed on 2008 January 24, from 02:07 to 03:00 UT when the observed radio wave was emitted on the Jupiter. The located DAM source corresponding to the frequency range of 4–10 MHz is shown in Figure 8, which the related location is between 0.4 and 0.9 $R_J$ above Jovian surface and M-shell value ranges from 6.6 to 16.6 $R_J$. The projected locations of source and Io (see Figure 9) show that the non-Io-A emission has no relation with Io, where averaged FP in longitude drifts from $161.2^{\circ}$ to $161.9^{\circ}$ with the range of $157^{\circ} - 164^{\circ}$ between 10 and 90 percent. Comparing the radio spectra in Figure 8 and projected location in Figure 9, the apparent rotation speed $\Omega_{DAM}$ is about 0.987 $\Omega_J$ with an error of $\pm 0.138\, \Omega_J$, and the rotation speed of source footprint $\Omega_{FP} \approx (0.951 \pm 0.034)\, \Omega_J$. The averaged emission angle (Figure 10) is about $63^{\circ}$ (quasi constant) and ranges from $56^{\circ}$ to $72^{\circ}$ between 10 and 90 percent of field lines. Correspondingly, the averaged energy is about 16.5 keV and ranges from 8 to 24 keV, which is much higher than the Io-C event on 2008 January 24.

## 3.4 Non-Io-C Event on 2009 August 13

The non-Io-C event was observed on 2009 August 13 from 01:50 to 03:00 UT without light travel time, corresponding to the fit for the arcs in frequency range of 6-11 MHz (see Figure 11). The related location is between 0.15 and 0.5 $R_J$ above Jovian surface and M-shell value ranges from 2.6 to 7.6 $R_J$. According to all the projected source location in Figure 12, the averaged FP in longitude drifts from 292° to 306° with the range of 290° - 308° between 10 and 90 percent, which is close to the Io-C event on 2008 January 24 but far away from Io footprint. The apparent rotation speed $\Omega_{DAM}$ is about 0.813 $\Omega_J$ with an error of $\pm$ 0.075 $\Omega_J$, which is larger than the rotation speed of FP $\Omega_{FP} \approx (0.664 \pm 0.005)\ \Omega_J$. Respectively, the averaged emission angle (Figure 13) has a decrease from $63^{\circ}$ to $62^{\circ}$ and the averaged energy increases from 12.1 to 12.9 keV which is higher than the Io-C event but lower than the non-Io-A event above.

# 4. Statistical Results

Not all the observed DAM arcs can find appropriate source regions through our method. In the 89 classified DAMs from multi-view spacecraft, there are 58 DAMs that we can find their sources based on at least two spacecraft observations, and all the other DAMs are excluded from the following statistical analysis. Plus the 10 evens from the NDA database, we have a total of 68 events (60 Io-DAMs and 8 non-Io-DAMs). Their inferred properties are listed in Table 2.

## 4.1 Reliability of the Method

To verify the reliability of the method, we compared the above results with previous studies. It is known that when Jupiter rotates, the magnetic field is dragged by the Io due to the slow rotation of the Io, while the Alfvén wave is excited by magnetic perturbance in the perpendicular direction, and then the magnetic field forms Alfvén wing extending to the polar region as it passes through the Io. The Io footprint (IFP) in the polar regions consists of a number of auroral spots (Gérard et al., 2006), with a longitude difference from the Io. The relationship between the longitude of Io, that of Io auroral spots and that of Io-DAM source footprints can therefore be established (e.g., Hinton et al., 2019, hereafter H19)

As the Figure 14(a) shows, the longitude difference between the footprints of Io-DAM and Io of our 68 events is plotted with the y-axis on the left. For comparison, the longitude difference between Io auroral spots and Io is superimposed with the y-axis on the right, which shows the prediction by Alfvén wing model (H19) and observation by HST (Bonfond et al., 2017, or B17). The source footprints of Io-DAM and the Io auroral points follow the same trend varying with Io's longitude, because they both depend on the magnetic field topology of Jupiter and the Io-Jupiter interaction. The footprints of active field lines are ahead of the Io auroral spots along the direction of Jupiter's rotation, related to the presence of the Alfvén wing.

Further, we calculate the lead angle, in which the Io footprint is obtained from the relationship between Io and the corresponding MAW UV spot established by Bonfond et al. (2009). As shown in Figure 14(b), the mean value of the lead angle ranges from 0° to 20°, with a variance similar to a sinusoidal function. Hess et al. (2010) fitted a cosine function of lead angle $7\cos(\Lambda_{Io} - 40) + 10$ (in degrees) in the MAW frame based on 32 Io-C and 20 Io-D events in the southern hemisphere. Both Io-C and Io-D in our results are consistent with this cosine function, where the Io-D event corresponds to a larger lead angle and the Io-C event corresponds to a smaller. All of these agreements suggest the reliability of our method.

## 4.2 Properties of the DAM Sources

The evolution of the FPs of active field lines on the surface of Jupiter is indicated by the arrow in Figure 15. The source locations of the DAMs are mainly concentrated in three regions. Two are in the southern hemisphere between about 50° and 130° for Io-D events and between about 250° and 350° for Io-C and non-Io-C events, the other in the northern hemisphere between about 150° and 200° (non-Io-A, Io-A and Io-B events). The latitudes of the three regions are mainly distributed around ± 60° near the Io footpath (green lines). In contrast to the CML-Io phase distribution studied previously, our study suggests that the actual source locations of A and B events in the northern hemisphere are concentrated and mixed together in the strong magnetic anomaly, while the source locations of C and D events in the southern hemisphere are separated though the magnetic field strength is almost uniform in the southern hemisphere.

The local times of these events are indicated by the blue-red colors. It is found that 8 non-Io events are all distributed on the dusk side, which is consistent with the previous results that the fast forward interplanetary shocks excite non-Io-DAMs almost exclusively on the dusk side (Hess et al., 2012, 2014). Besides, the active field lines of Io-DAM emissions span more widely in longitude than non-Io-DAM events. This might be because the active field lines of non-Io events co-rotate with Jupiter, and their observed duration is shorter compared to Io events (roughly co-rotating with Io) for a given longitude difference of the multi-view observers. Similarly, for the NDA events, since they are studied based on single-point observations, their evolution times are negligible in Fig.15.

The results for the distribution of source locations are independent of the observed time, which is spanning from 2008 to 2014 September, and independent of the variation in observer's latitude. Different from remote sensing observations, Louis et al. (2019) analyzed the source locations of Jupiter radio emissions in Juno's first 15 perijoves by using in situ observations Juno/Waves. The method to identify the source location is comparing the local cyclotron frequency with the observed emission frequency. Their result showed that the source locations of 28 non-Io-DAM events in both the northern and southern hemispheres almost scatter in the whole longitude range. This is different from the source region distribution of our 8 non-Io events in Figure 15, which may be due to the bias of the selected events, which is toward the intense and long-lasting DAMs. The events selected in our study are of high radiation intensity and long duration (more than 30 min) based on remote sensing observations, while Louis et al. (2019) observed one moment of the non-Io-DAM source (2-3 min on average) with a small segment frequency due to the Juno orbit limitation, as well as being able to observe many events of weaker intensity.

In addition to the source location distribution, the emission angle and corresponding electron energy for different types of the DAM are also inferred. Figure 16 shows the average of emission angle and resonant energy for all fitted events with the range from 10 to 90 percent (gray bars). The emission angle for Io-A events is distributed on average from 65° to 75°, which is close to the range of 62° to 72° for Io-B. The emission angle of Io-D is generally smaller, with mean value distributed between 60° and 65°, while the emission angle of Io-C is the largest, with mean value distributed between 75° and 85°. The non-Io events have smaller emission angle than Io-DAM events for the same type. Correspondingly, the DAM electron energy is mainly distributed between 0.5 and 20 keV, the non-Io-A events have the largest electron energy with the main range of 15 to 20 keV, and the Io-C events have the smallest electron energy with the range of 0.5 to 5 keV.

## 5. Summary and Discussion

In this paper, we present the statistical study of DAM events based on multi-view observation with Wind and STEREO spacecraft. We select 111 events in 2008-2014 September and provide a statistical standard to classify Io-DAM and non-Io-DAM events by apparent rotational speed in radio spectra, where the rotational speed of Io-DAM is in range of 0.15-0.6 $\Omega_J$ and that of non-Io-DAM is between 0.7-1.2 $\Omega_J$. Based on CML-Io phase diagram and the tilt of Jovian magnetic dipole axis, we classified the events, and identified 15 Io-C, 60 Io-D, 3 non-Io-A, 10 non-Io-C and one non-Io-D.

Further, we successfully inferred the source location, active field lines, emission angle and electron energy of 68 events (including 10 NDA events) using the method by Wang et al. (2020). The distribution of different types of DAM source regions shows a north-south asymmetry, with one concentrated source region in the northern hemisphere around 150° to 200°, i.e., the magnetic anomaly and two separated source regions in the southern hemisphere around 50° to 130° and around 250° to 350°, which may a result of the magnetic field topology of the Jovian magnetospher. And each type of event has a specific preferred range of source regions. We compared our results with Louis et al. (2019) results based on in-situ observations from Juno, and suggested that the intense and long-lasting non-Io DAMs have preferred source regions though weak and short non-Io DAMs could be excited at any longitude.

**Acknowledgments.** We acknowledge the use of the data from the radio instruments on board Wind, STEREO-A and B spacecraft and the DAM events from Nancay Decametric Array. The work done by the authors in China is support by the Strategic Priority Program of the Chinese Academy of Sciences (Grant No. XDB41000000) and the NSFC (Grant Nos 42188101 and 42130204).

# Figures and Tables

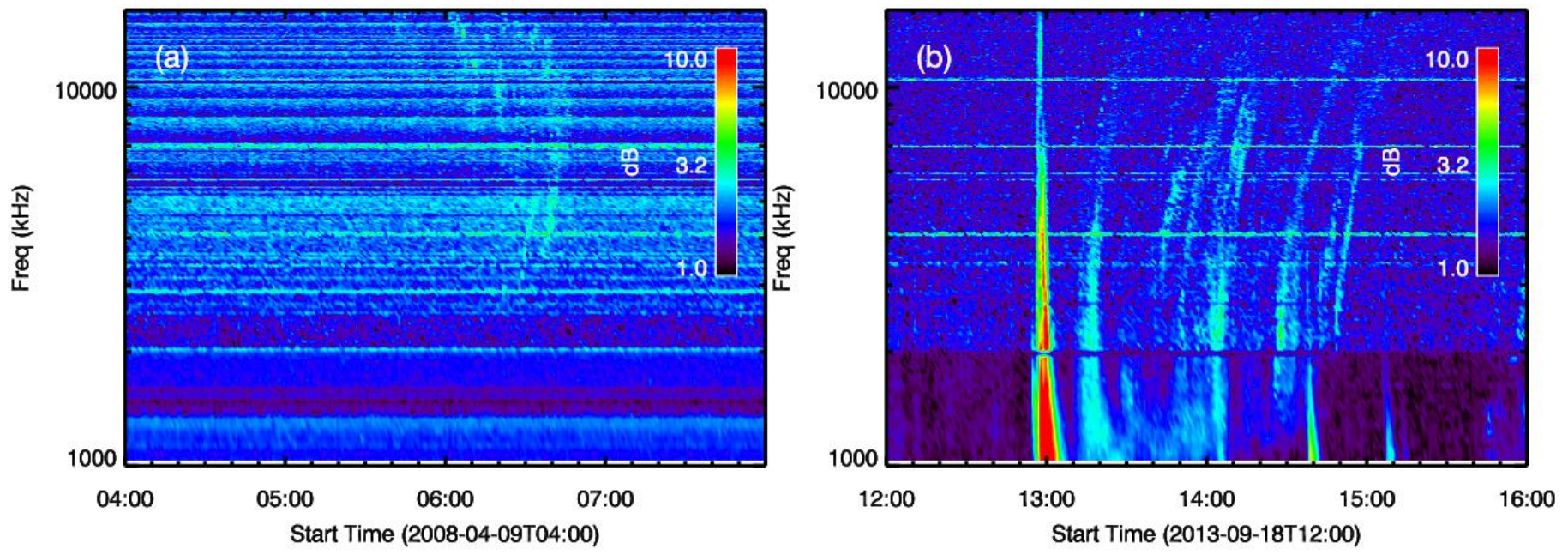

**Figure 1.** Examples of excluded events with complex structures. Panel (a) shows the ambiguous boundaries of the arcs and panel (b) shows a group of arcs with tens of minutes intervals in the spectrum.

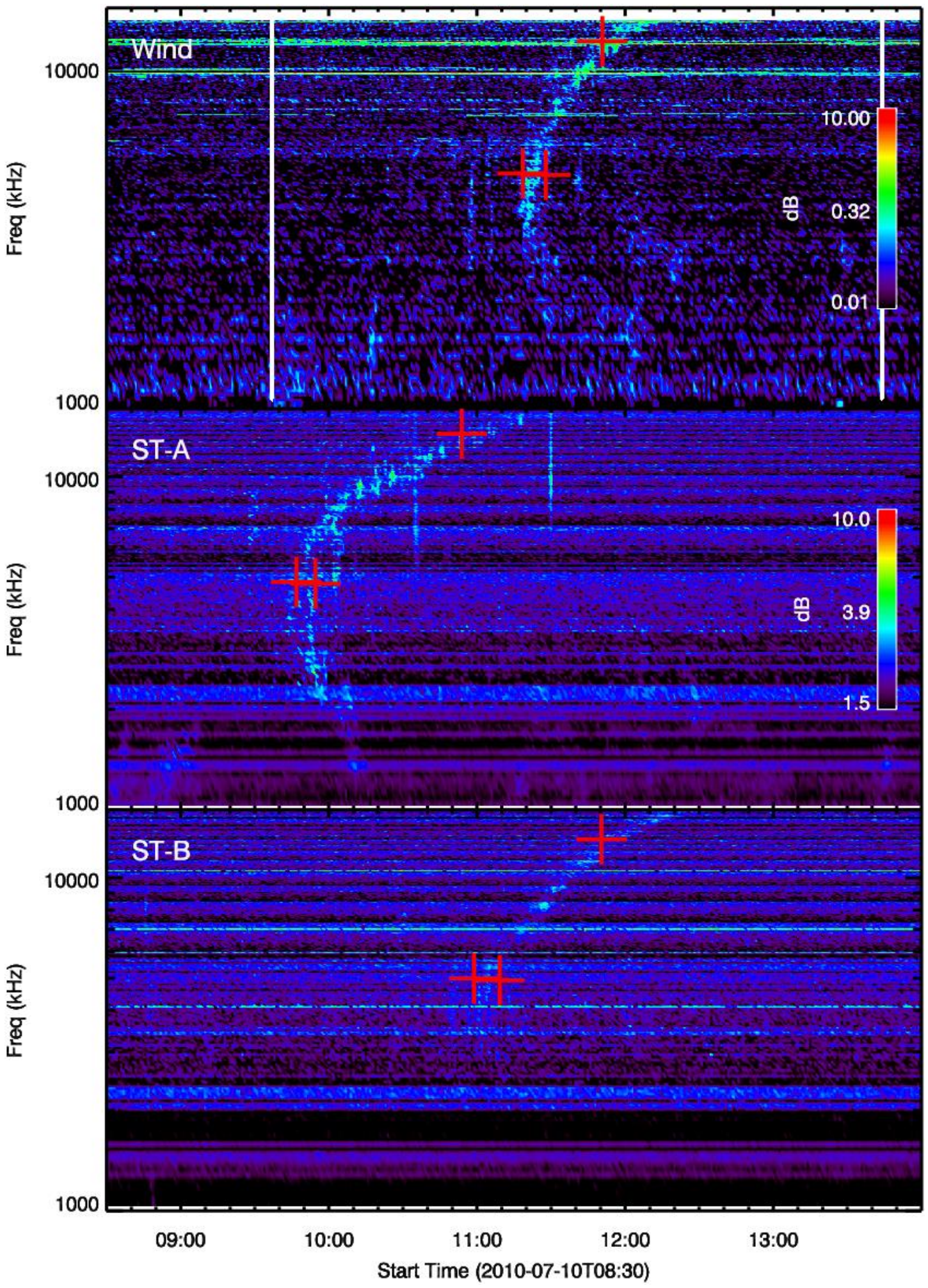

**Figure 2.** Radio dynamic spectrum from Wind/WAVES, ST-A/WAVES, ST-B/WAVES. The red markers indicate the start and end time at the apex of the arc and the start time of the leading end of the arc.

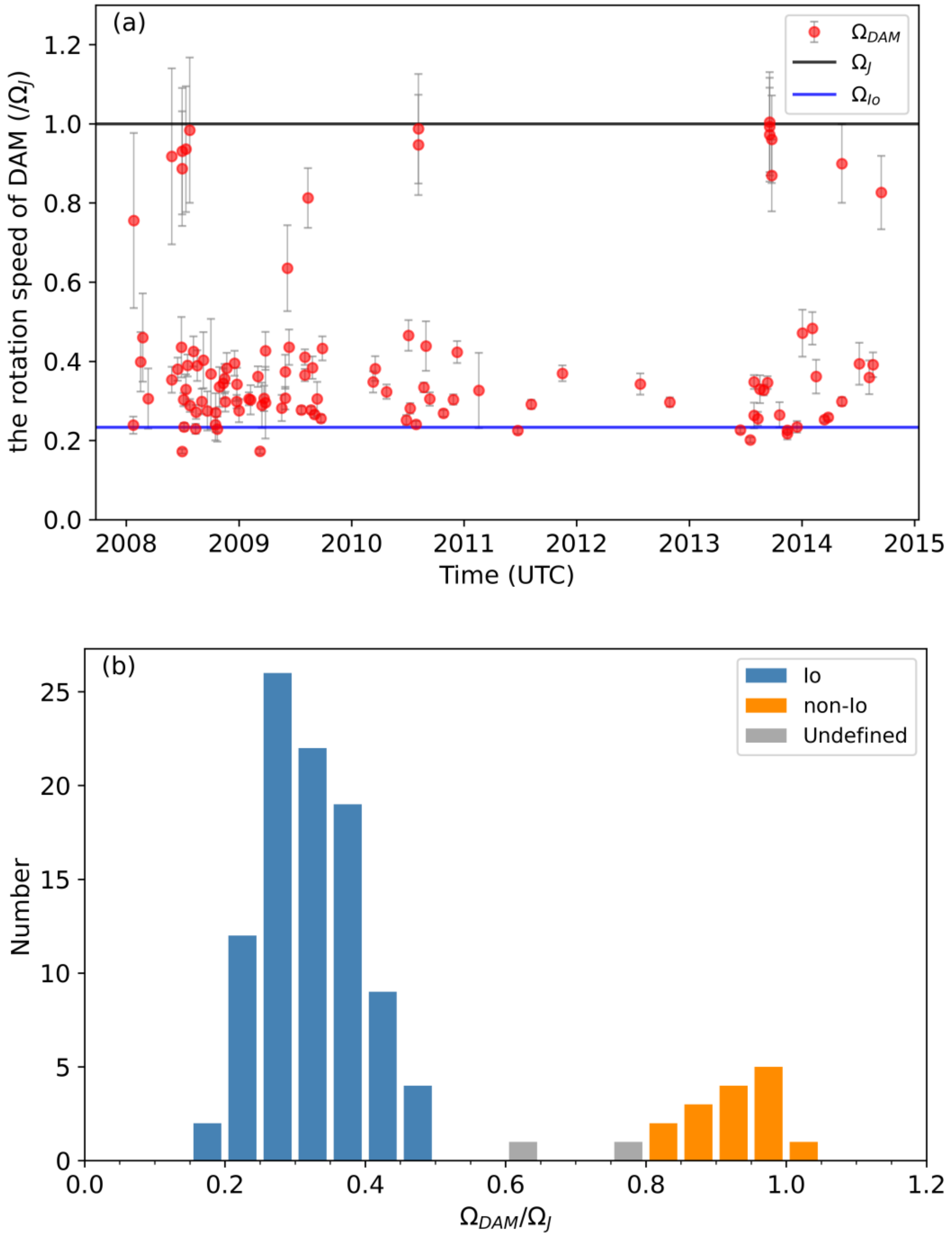

**Figure 3**. Panel (a) shows the statistical rotation speed of DAM emissions with time. The red circles give the median values and error bars indicate the time error of considering ±4 min on the radio dynamic spectrum. The rotational angular speed of Io and Jupiter are represented by blue line and black one respectively. Panel (b) is the histogram of the rotational angular speed of DAM emissions. These events can be divided into Io-related (blue bars), non-Io (orange bars) DAM and undefined events (gray bars) with error.

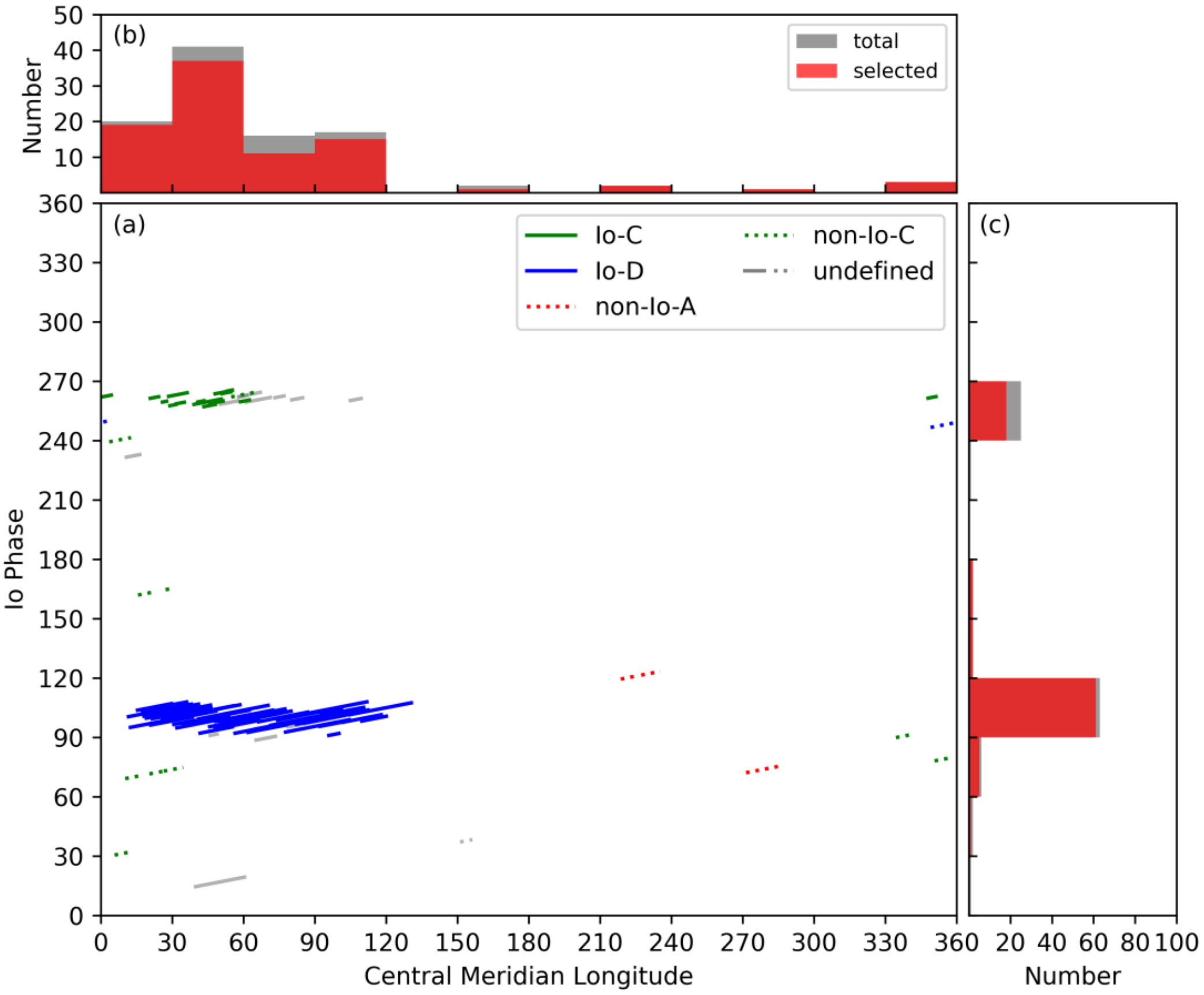

**Figure 4**. Io phase versus west system III longitude of Wind. Panel (a) represents the CML and Io phase ranges over the entire Wind observation period of DAM emission, in which solid line means Io-DAM and dash line means non-Io-DAM. The colored part is the events selected for simulation and judged to be in the four source areas A, B, C, and D, while the gray part indicates unselected events. Panel (b) and (c) are histograms of CML and Io phase in 30° bins, which red bars mean the colored events in panel (a).

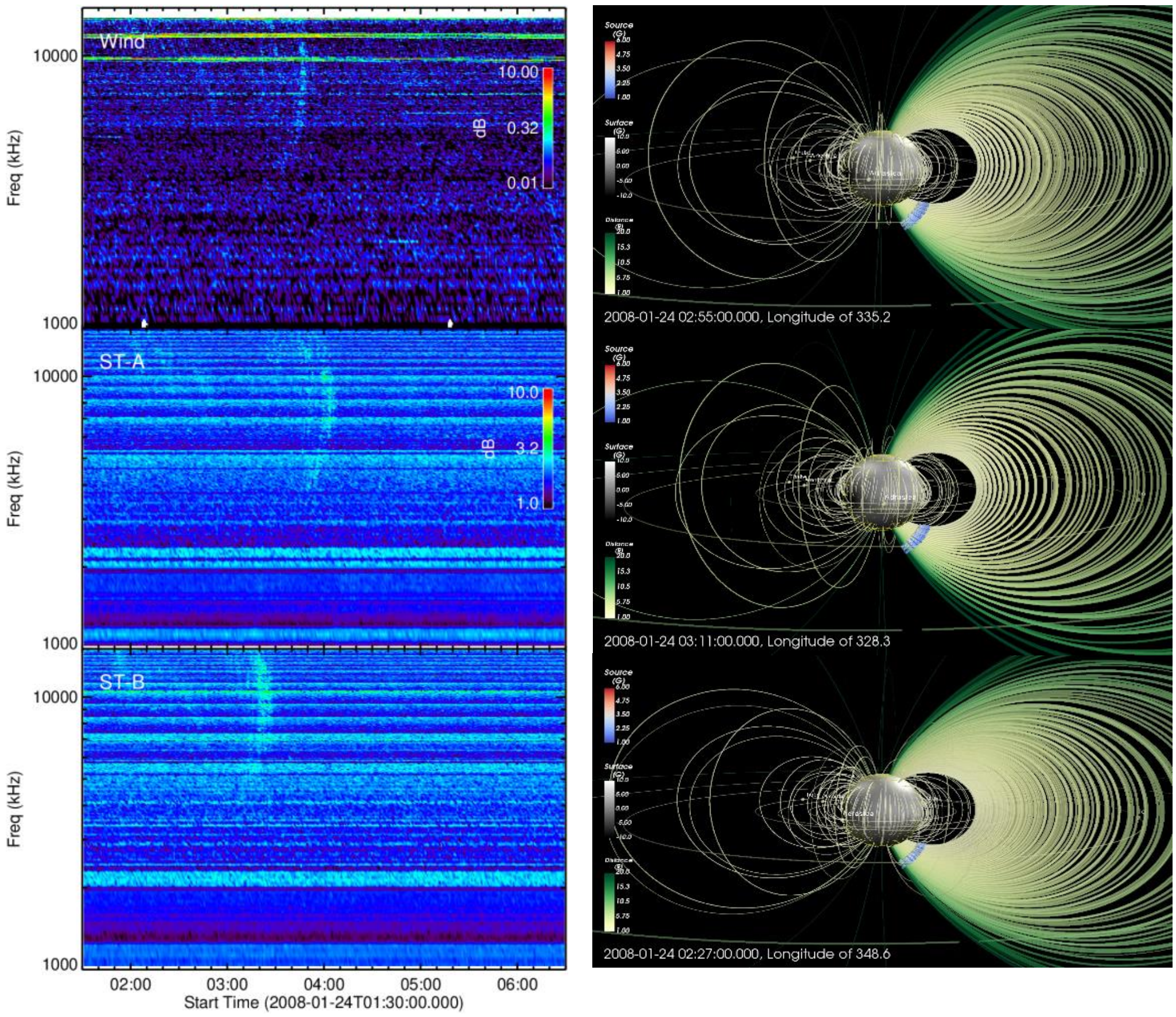

**Figure 5**. Observation and model of the Io-C emission on 2008 January 24. (left panel) Radio dynamic spectra from Wind, ST-A and ST-B, respectively. (right panel) The DAM source location by model at the first view of related observers, in which yellow-to-green solid lines represent active field lines and blue-to-red dots represent located source.

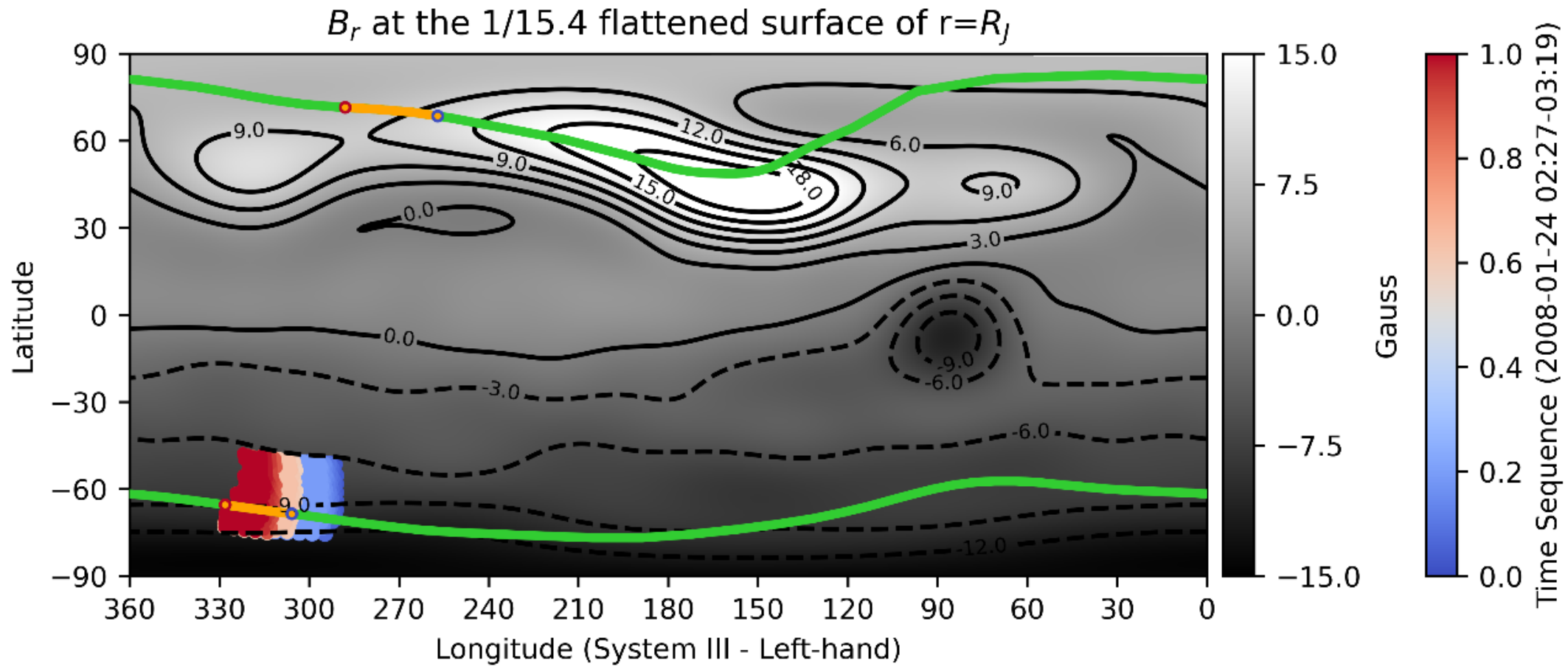

**Figure 6**. The projected location of DAM at flattened Jovian surface of 1 $R_j$. All modeled source footprints are shown as blue-to-red dots varying with observed time. The Io footprints are shown as green lines in the northern and southern hemisphere according to the where orange segments represent the local Io footprints at observation time.

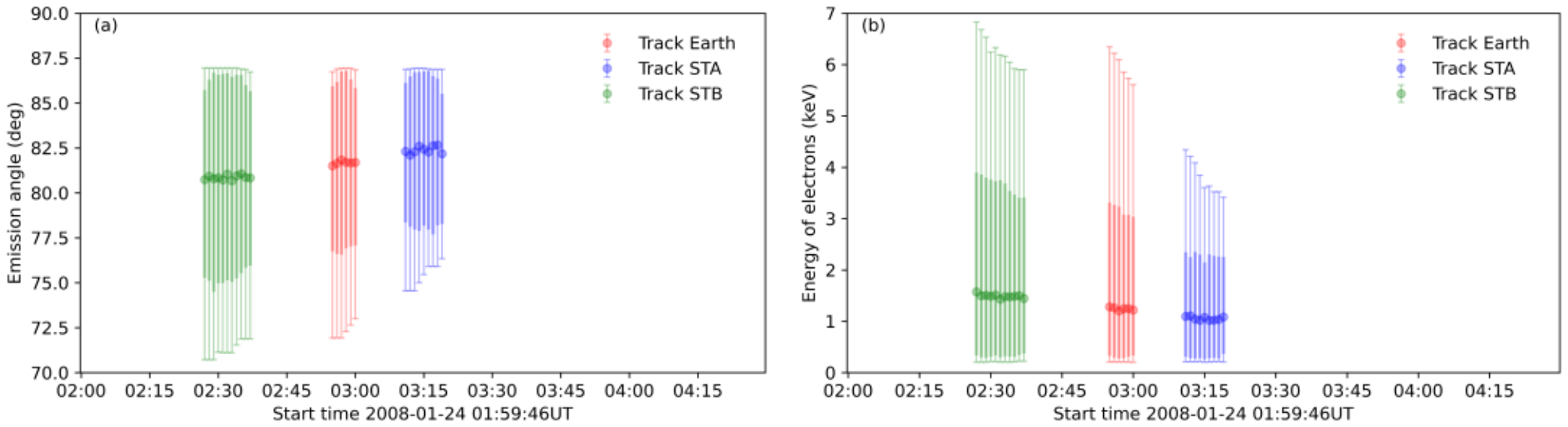

**Figure 7**. (a) Emission angle and (b) Energy of active electron during the Io-C emission on 2008 January 24. The circles give the median values, the thin bars give the minimum and maximum values and the thick bars give the top and bottom 10th percentiles.

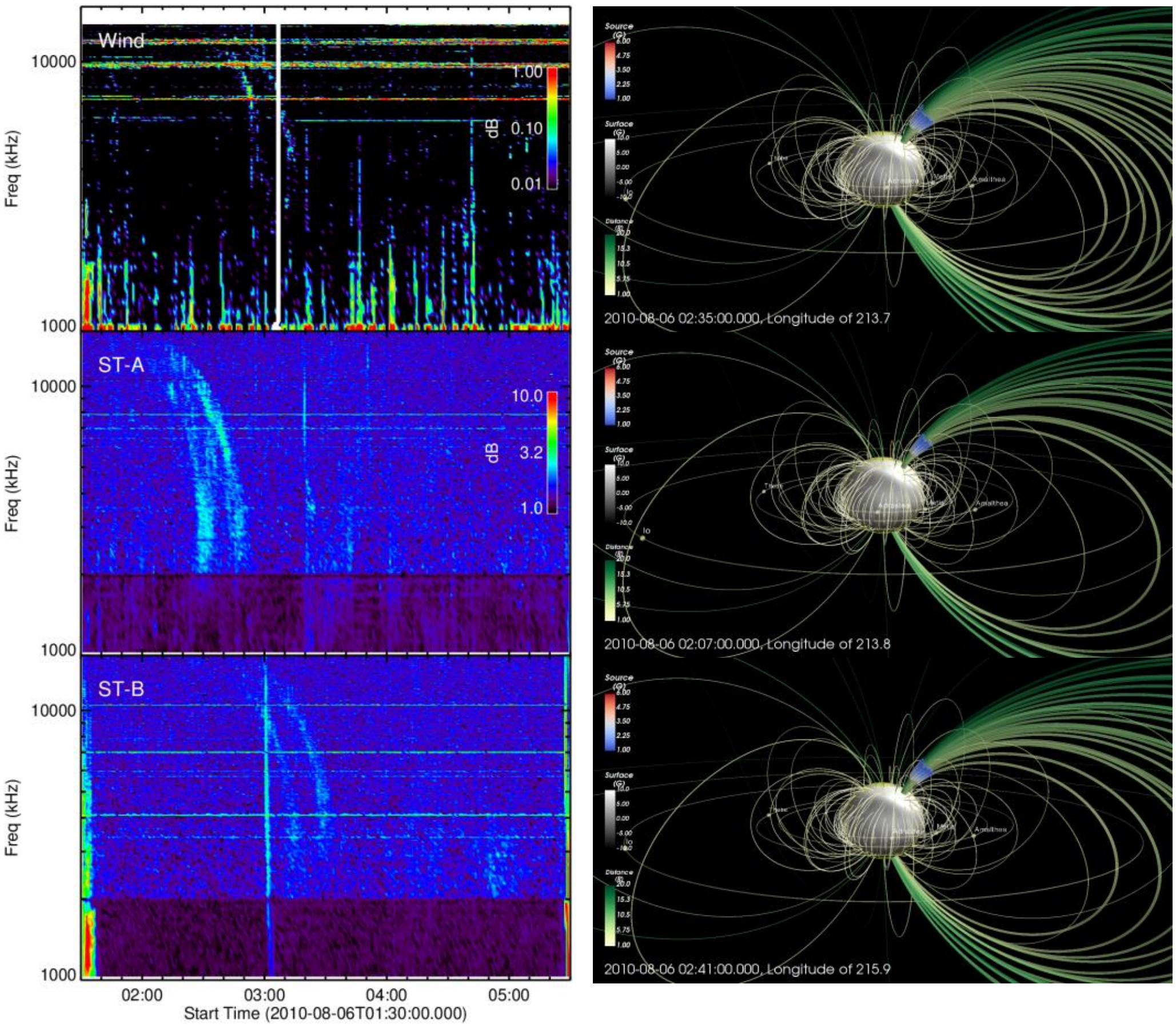

**Figure 8**. As Figure 5, but for the non-Io-A emission on 2010 August 6.

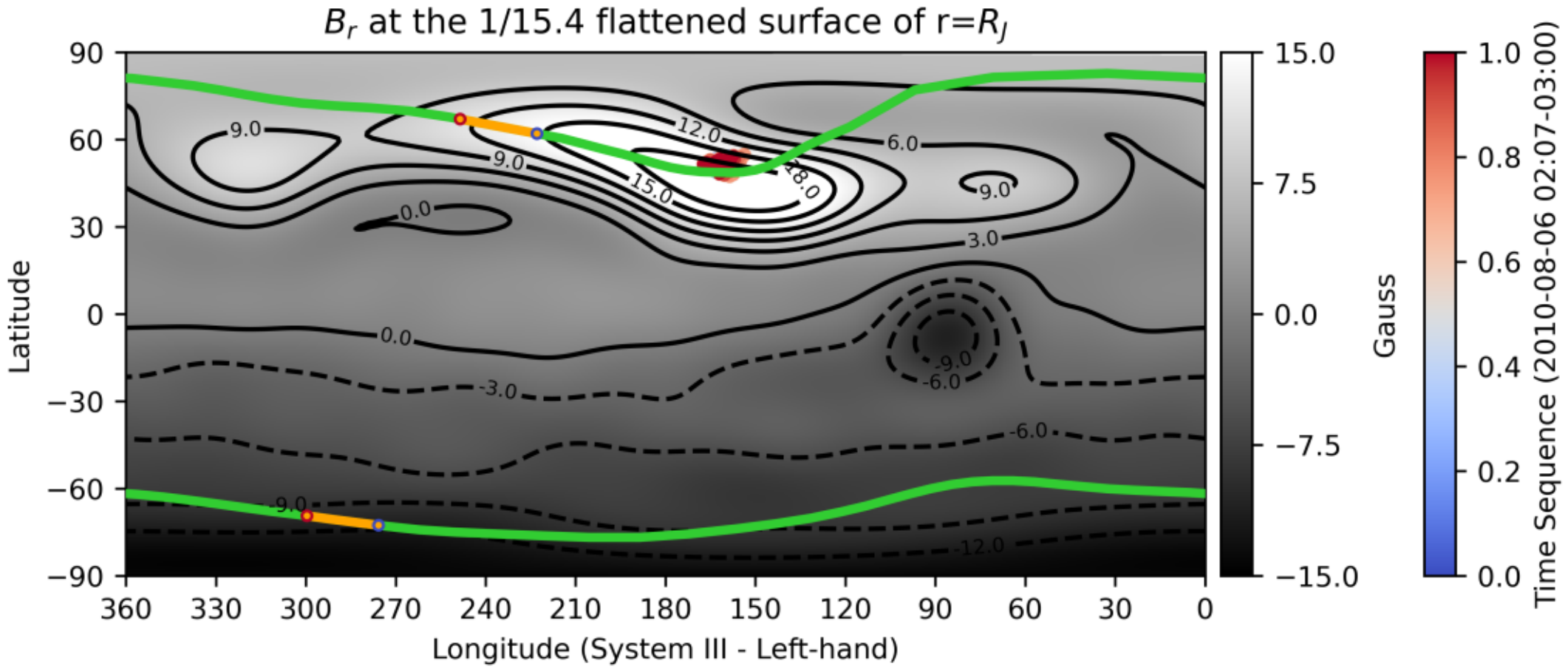

**Figure 9**. As Figure 6, but for the non-Io-A emission on 2010 August 6.

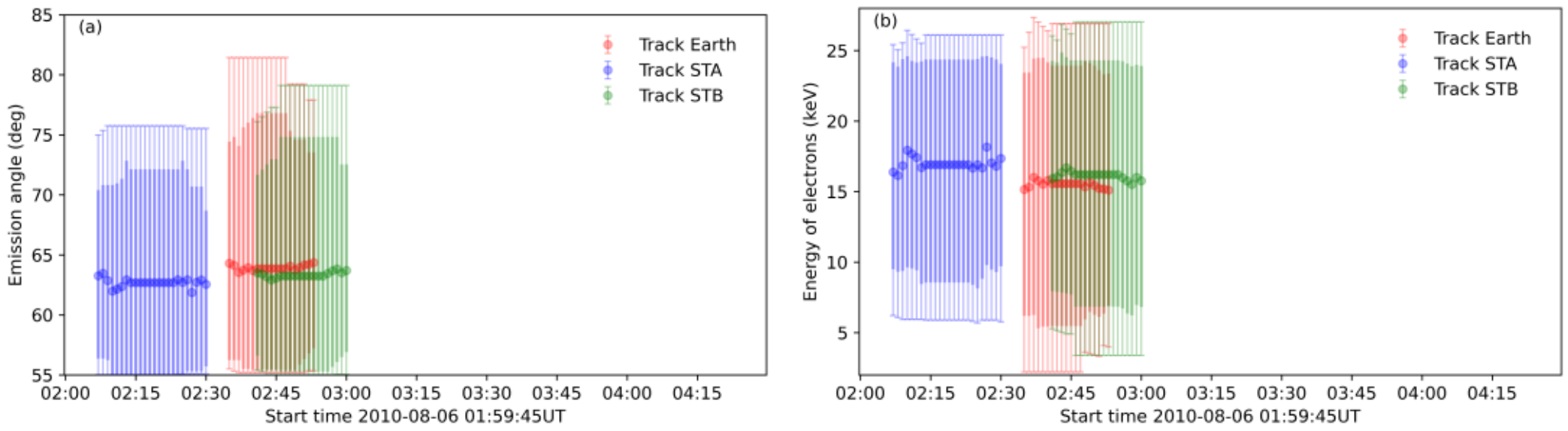

**Figure 10**. As Figure 7, but for the non-Io-A emission on 2010 August 6.

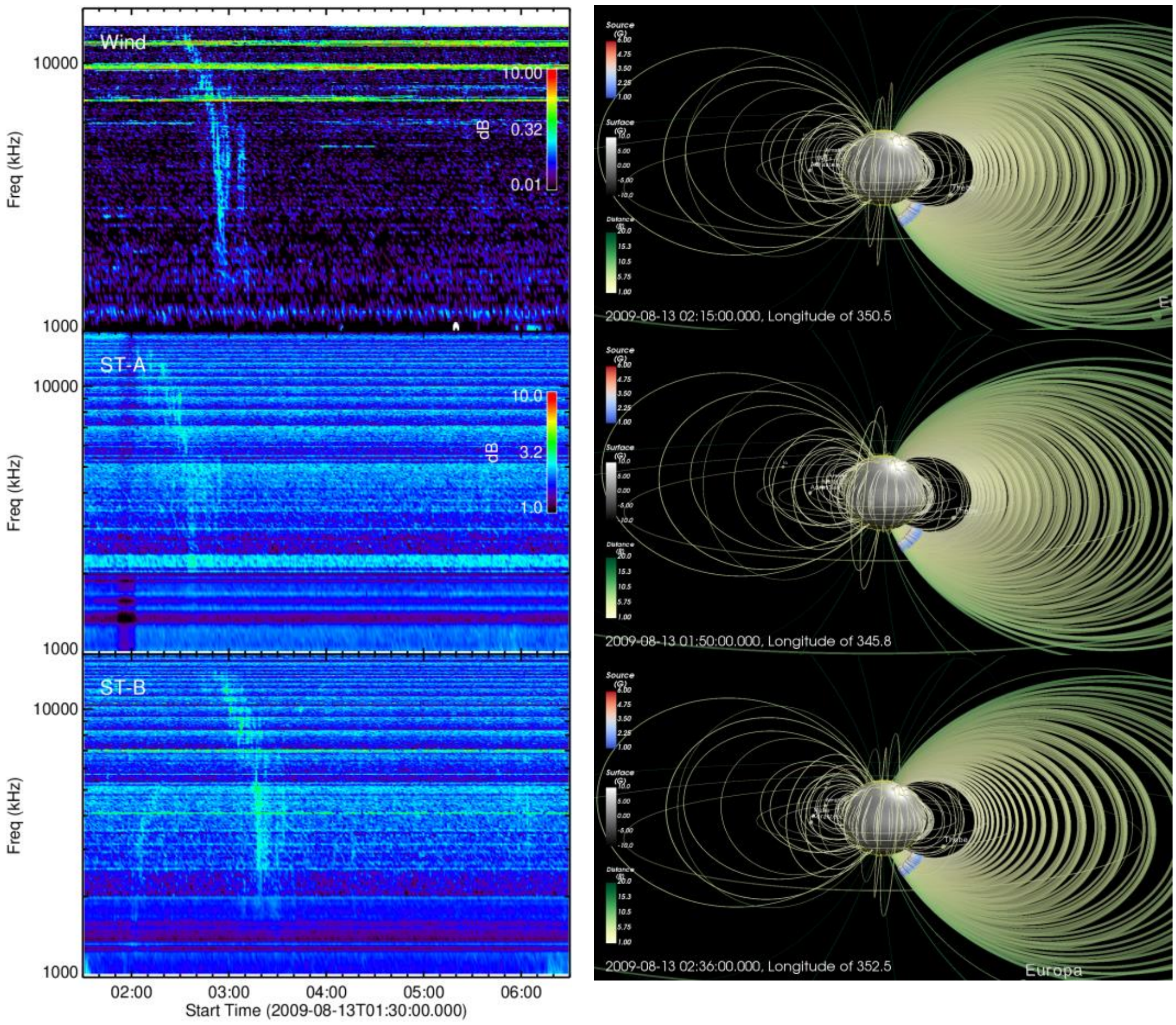

**Figure 11**. As Figure 5, but for the non-Io-C emission on 2009 August 13.

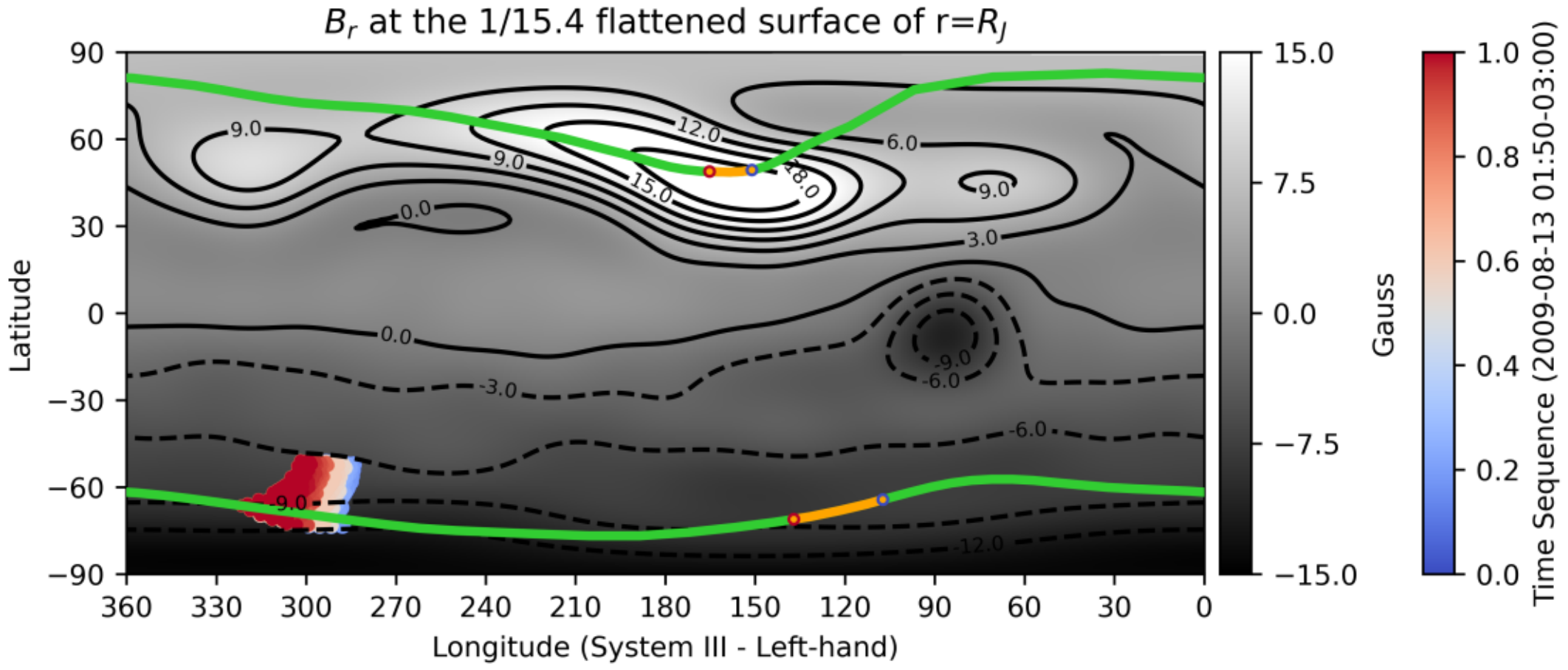

**Figure 12**. As Figure 6, but for the non-Io-C emission on 2009 August 13.

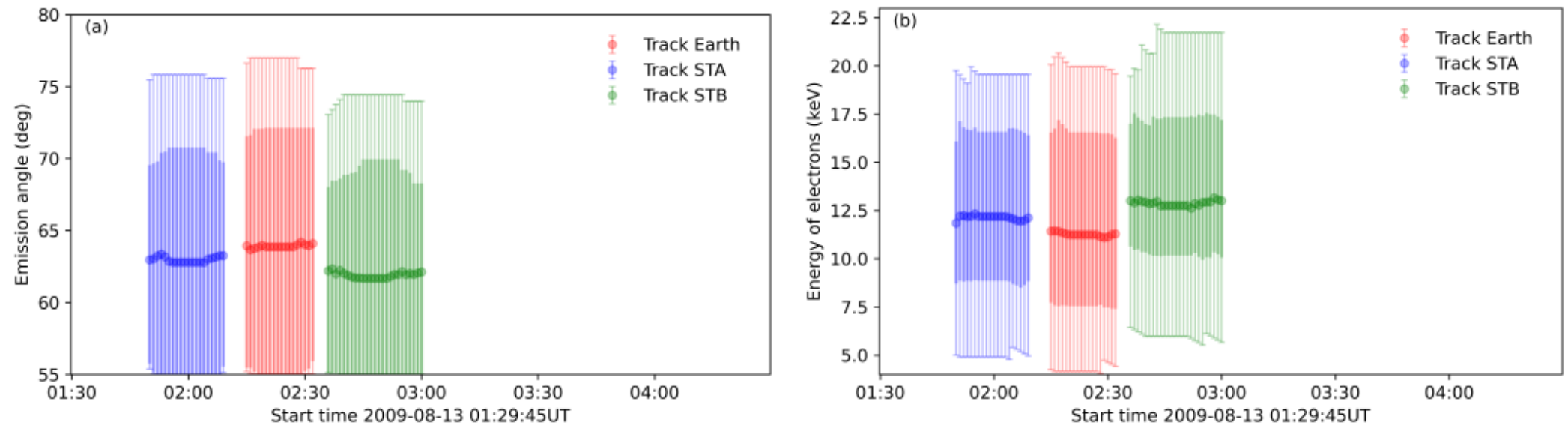

**Figure 13**. As Figure 7, but for the non-Io-C emission on 2009 August 13.

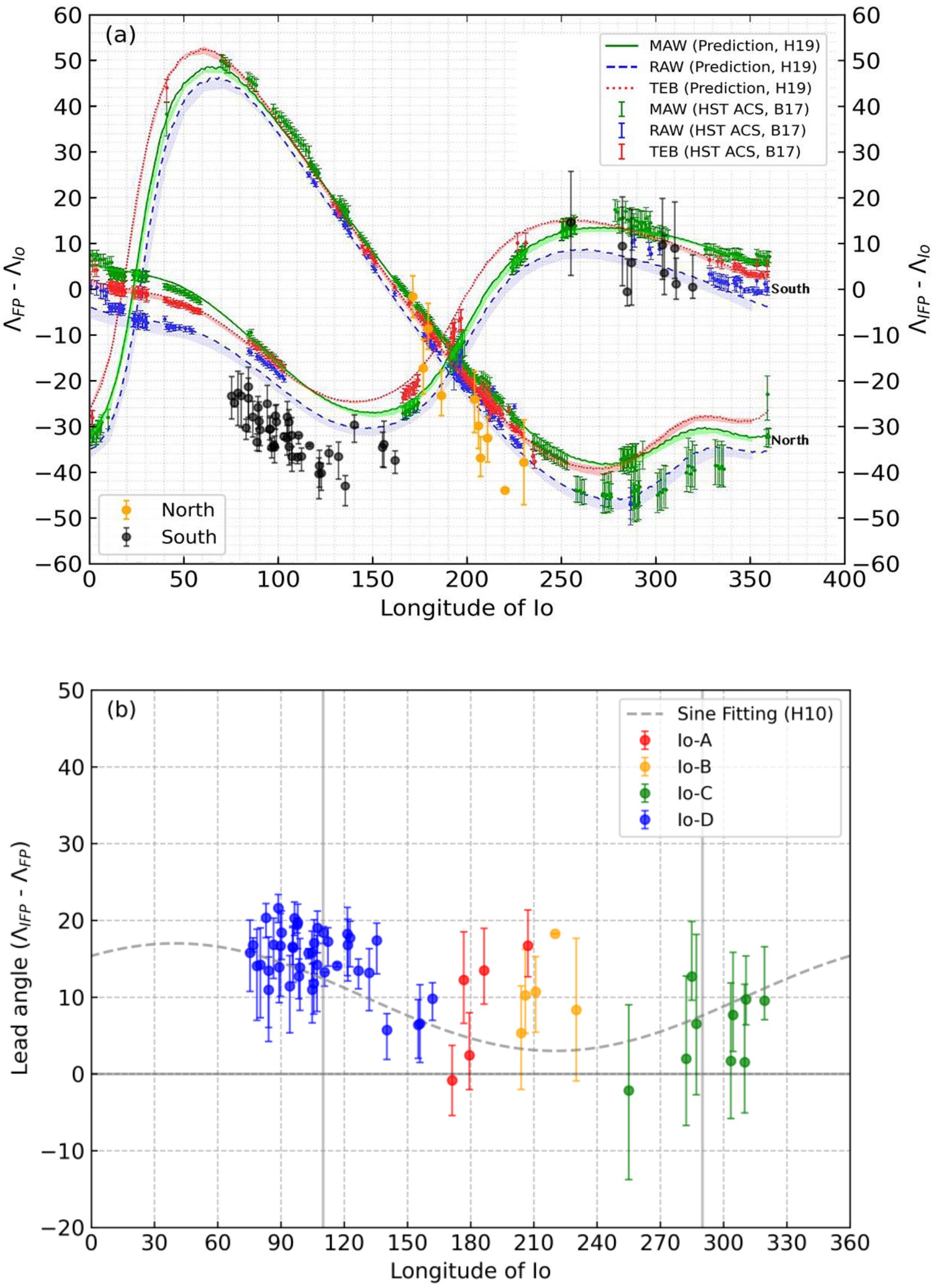

**Figure 14**. Longitude difference between Io, Io auroral spots and source footprints. Panel (a) shows longitude difference between Io and simulated footprints of Io-DAM source for the lower left legend and upper right legend is cited from Hinton et al. (2019, hereafter H19), predicting main Alfvén wing (MAW), transhemispheric electron beam (TEB) and reflected Alfvén wing (RAW) for all longitude of Io with HST observation of auroral spots (Bonfond et al., 2017, hereafter B17). Panel (b) shows the projected lead angle between IFP at MAW spots frame and FPs with the fitting sinusoidal function cited from Hess et al. (2010, hereafter H10).

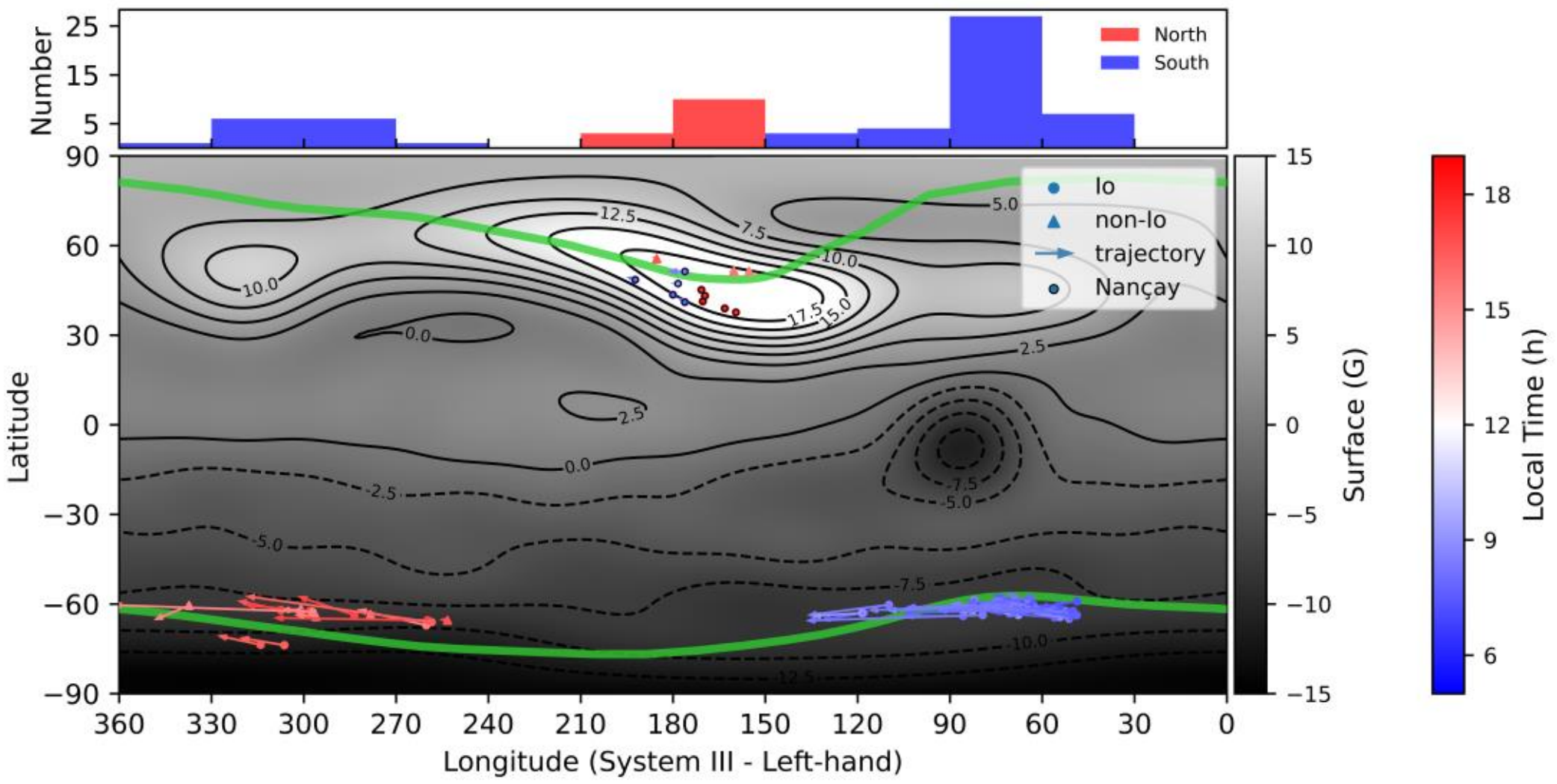

**Figure 15**. The projected location of DAM source. The Jupiter surface is flattened by a factor of 1/15.4 at 1 $R_j$, and the gray-scale values in background denote the magnetic field strength of $B_r$. The color-coded arrows indicate the trajectory of the footprints of active field lines corresponding to DAM source region, where the dotted and triangular markers show the starting footprints for Io and non-Io events, respectively. The blue-to-red color indicates the longitude difference from the Sun and source footprint, with negative values on the dawn side and positive values on the dusk side at the right color-bar. The histogram at the top indicates the distribution of longitude averages in the source region in 30° bins.

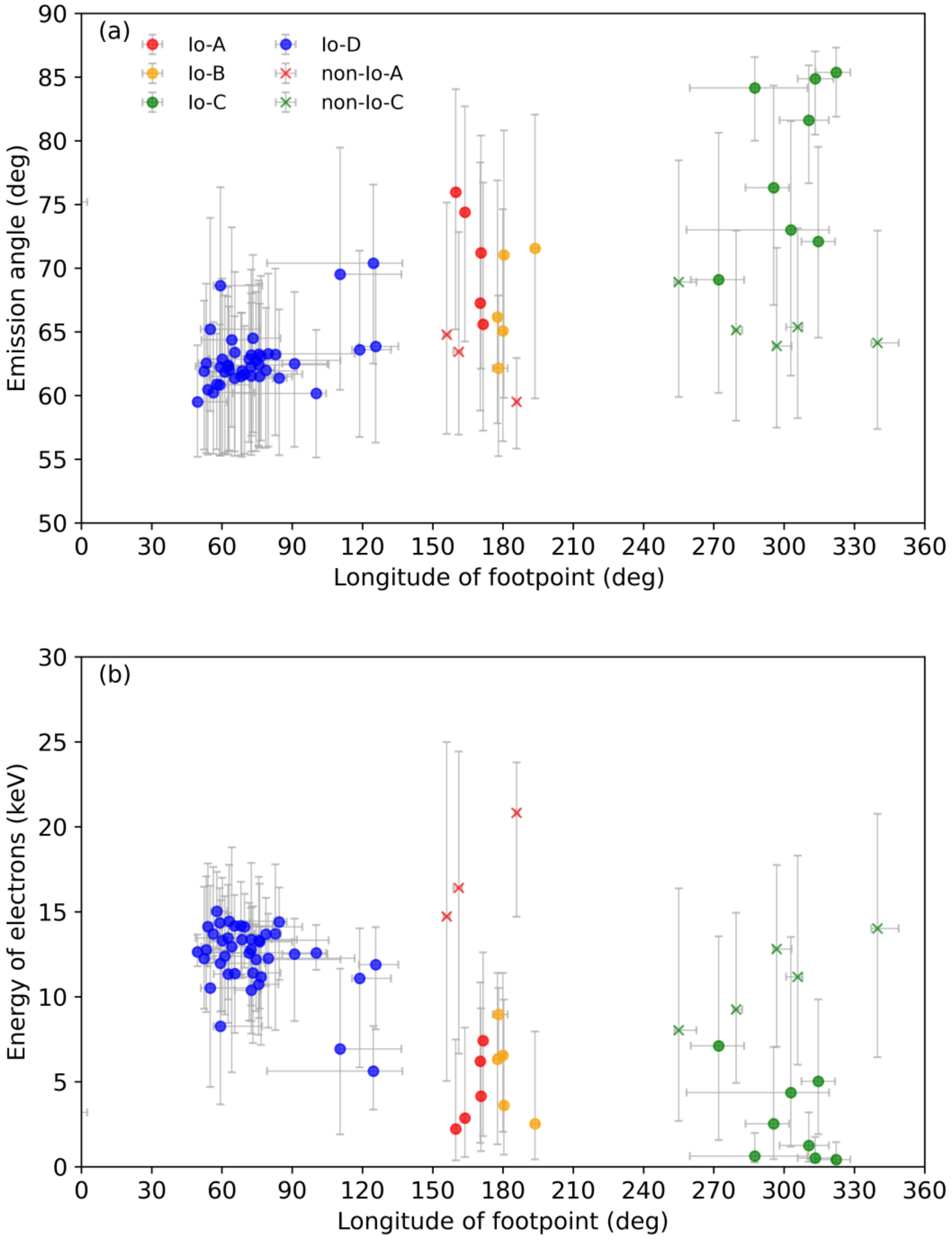

**Figure 16**. (a) Emission angle and (b) Active electron energy versus longitude of DAM source footprints for different types (colored markers).

**Table 1.** Statistics of emission parameters for observations.

| No. | Date | Wind | | | | | Stereo-A | | | | | Stereo-B | | | | | VE/VL | Type | Rotation Speed [$\Omega_J$] |
|---|---|---|---|---|---|---|---|---|---|---|---|---|---|---|---|---|---|---|---|
| | | Apex time [UT] | Arc width [min] | Span [min] | Apex freq. [kHz] | End freq. [kHz] | Apex time [UT] | Arc width [min] | Span [min] | Apex freq. [kHz] | End freq. [kHz] | Apex time [UT] | Arc width [min] | Span [min] | Apex freq. [kHz] | End freq. [kHz] | | | |
| (1) | (2) | (3) | (4) | (5) | (6) | (7) | (8) | (9) | (10) | (11) | (12) | (13) | (14) | (15) | (16) | (17) | (18) | (19) | (20) |
| 1* | 2008-01-24 | 03:45:31 | 4.3 | 7.7 | 8860 | 5420 | 03:59:38 | 6.7 | 10.1 | 8970 | 5280 | 03:18:51 | 8.9 | 11.6 | 8870 | 6090 | VL | Io-C | 0.24 |
| 2 | 2008-01-25 | 18:24:54 | 3.6 | 15.3 | 4000 | 9940 | 18:27:38 | 4.6 | 18.9 | 3920 | 9800 | 18:16:58 | 6.3 | 16.4 | 3960 | 10240 | VL | Undefined | 0.76 |
| 3* | 2008-02-17 | 02:51:50 | 8.6 | 5.4 | 4040 | 8070 | 02:53:45 | 9.5 | 0.3 | 3920 | 8770 | 02:37:12 | 6.4 | 22 | 4000 | 9800 | VE | Io-D | 0.40 |
| 4* | 2008-02-24 | 04:20:04 | 6.9 | 15.4 | 3960 | 10050 | 04:23:41 | 8.4 | 13.3 | 3920 | 10020 | 04:12:14 | 6 | 5.7 | 3870 | 9170 | VE | Io-D | 0.46 |
| 5 | 2008-03-12 | 22:09:32 | 5.7 | 23.6 | 4830 | 11390 | 22:07:20 | 8.2 | 30.8 | 4780 | 12500 | 21:57:17 | 10 | 35.2 | 4940 | 12780 | VE | Io-D | 0.31 |
| 6* | 2008-05-27 | 23:23:14 | 4.5 | 13.9 | 4880 | 9740 | 23:10:11 | 4.8 | 13.6 | 4880 | 11570 | 23:31:10 | 5.7 | 13.3 | 4830 | 11310 | VL | Non-Io-C | 0.92 |
| 7 | 2008-05-28 | 00:38:07 | 8.2 | 14.3 | 4980 | 12130 | 00:09:50 | 8.5 | 13 | 5050 | 11820 | 00:57:12 | 5.4 | 10.6 | 4830 | 11820 | VE | Io-D | 0.35 |
| 8 | 2008-06-16 | 10:42:17 | 7.6 | 29.2 | 3870 | 11280 | 10:16:29 | 8.7 | 33.6 | 4000 | 12400 | 11:10:45 | 8.9 | 19.7 | 3860 | 13060 | VE | Io-D | 0.38 |
| 9* | 2008-06-28 | 20:23:26 | 9 | 14.1 | 3990 | 10950 | - | - | - | - | - | 20:47:45 | 8 | 20.9 | 4020 | 11880 | VE | Io-D | 0.44 |
| 10 | 2008-06-30 | 13:54:24 | 12 | 32.2 | 5040 | 10700 | 12:47:52 | 10.3 | 18.5 | 4880 | 12360 | 14:54:04 | 9.4 | 9.9 | 4830 | 12640 | VE | Io-D | 0.17 |
| 11 | 2008-06-30 | 22:19:14 | 5.4 | 17.5 | 3420 | 8320 | 22:06:47 | 6.8 | 12.4 | 3430 | 7030 | 22:31:23 | 5.4 | 22.6 | 3350 | 9910 | VL | Non-Io-C | 0.89 |
| 12 | 2008-06-30 | 22:31:57 | 5.1 | 10.8 | 2920 | 6200 | 22:21:20 | 3.8 | 10.2 | 2920 | 5990 | 22:44:46 | 3.4 | 9.5 | 3000 | 5710 | VL | Non-Io-C | 0.93 |
| 13* | 2008-07-05 | 21:43:18 | 7.8 | 18.4 | 3910 | 11280 | 20:56:50 | 9.8 | 32.2 | 4000 | 11520 | 22:08:47 | 4.7 | 7.1 | 3900 | 8580 | VE | Io-D | 0.30 |
| 14* | 2008-07-06 | 17:17:08 | 6.4 | 12.2 | 8810 | 4860 | 16:27:46 | 5 | 9.6 | 8870 | 4830 | 18:01:03 | 7.6 | 11.7 | 9040 | 4520 | VL | Io-C | 0.23 |
| 15* | 2008-07-12 | 21:23:37 | 3 | 13.9 | 7820 | 11620 | 21:13:53 | 3.3 | 14.6 | 7740 | 12530 | 21:35:38 | 3 | 14.9 | 7880 | 13060 | VL | Non-Io-C | 0.94 |
| 16 | 2008-07-12 | 22:44:12 | 8.3 | 3.6 | 4030 | 6350 | 22:20:25 | 12.6 | 7.9 | 3960 | 10060 | 23:19:10 | 7.9 | 16.8 | 3900 | 11630 | VE | Undefined | 0.33 |
| 17* | 2008-07-18 | 07:05:44 | 5.8 | 13 | 4980 | 10050 | 06:34:38 | 9.2 | 24.2 | 4940 | 13650 | 07:27:59 | 7.3 | 35.7 | 4670 | 12780 | VE | Io-D | 0.39 |
| 18 | 2008-07-25 | 01:33:29 | 5 | 13.6 | 2980 | 6210 | 01:26:34 | 6.4 | 13.5 | 3000 | 6370 | 01:44:16 | 6.4 | 15.2 | 2910 | 6730 | VL | Undefined | 0.98 |
| 19 | 2008-07-26 | 03:46:55 | 6.1 | 8.9 | 9940 | 4090 | 03:20:12 | 9.6 | 13.6 | 9910 | 4470 | 04:28:59 | 8.6 | 9.6 | 9690 | 4000 | VL | Io-C | 0.29 |
| 20* | 2008-08-06 | 17:27:43 | 8.3 | 20.6 | 4990 | 10510 | 17:16:55 | 9.1 | 19.4 | 4920 | 10430 | 17:55:44 | 4.4 | 7.4 | 5010 | 10760 | VE | Io-D | 0.42 |
| 21* | 2008-08-13 | 18:44:30 | 7.4 | 15 | 4910 | 9440 | 18:39:45 | 5.4 | 19.9 | 4940 | 11890 | 19:48:00 | 5.4 | 7.7 | 4860 | 9630 | VE | Io-D | 0.23 |
| 22 | 2008-08-14 | 14:38:05 | 4.4 | 8.6 | 7020 | 11780 | 14:22:49 | 4 | 9.8 | 6820 | 12680 | 15:18:13 | 3.8 | 10.6 | 6950 | 12810 | VL | Io-C | 0.27 |
| 23* | 2008-08-19 | 03:06:16 | 7.6 | 9.9 | 6020 | 10590 | 03:03:04 | 6.1 | 21.4 | 6030 | 12360 | 03:35:55 | 9.9 | 16.6 | 6090 | 12640 | VE | Io-D | 0.39 |
| 24* | 2008-09-02 | 05:44:35 | 6 | 20.1 | 5530 | 11760 | 05:43:18 | 6.4 | 20.7 | 5400 | 11820 | 06:11:41 | 5.7 | 14.7 | 5520 | 12220 | VE | Io-D | 0.30 |
| 25* | 2008-09-07 | 13:58:05 | 7.6 | 17.6 | 7670 | 13490 | 14:01:41 | 3.6 | 26 | 8070 | 13620 | 14:15:09 | 7.4 | 32.6 | 7970 | 14350 | VE | Io-D | 0.40 |
| 26* | 2008-09-20 | 18:25:24 | 7.2 | 12.6 | 6680 | 12390 | 18:37:26 | 6.6 | 13.2 | 6880 | 11820 | 18:42:53 | 7.8 | 22.4 | 6960 | 13800 | VL | Io-C | 0.27 |
| 27* | 2008-10-02 | 09:00:16 | 6.6 | 16.9 | 5980 | 12340 | 09:13:34 | 6.8 | 27.4 | 5960 | 13070 | 09:05:35 | 3.6 | 30.2 | 5870 | 12920 | VE | Io-D | 0.37 |
| 28* | 2008-10-16 | 12:37:12 | 4 | 10.2 | 6950 | 11900 | 13:06:23 | 6.2 | 10.8 | 6820 | 12550 | 12:35:52 | 4.8 | 23.5 | 7100 | 13240 | VE | Io-D | 0.24 |
| 29 | 2008-10-17 | 07:36:26 | 5.2 | 10.8 | 7820 | 4880 | 08:03:15 | 5.8 | 11.4 | 7940 | 4050 | 07:31:47 | 5 | 14 | 7940 | 4180 | VL | Undefined | 0.27 |
| 30 | 2008-10-22 | 15:04:41 | 5.9 | 12.4 | 10050 | 6270 | 15:38:24 | 8.7 | 18.6 | 10020 | 4140 | - | - | - | - | - | VL | Io-C | 0.23 |
| 31* | 2008-10-29 | 17:12:51 | 6.4 | 12.6 | 8770 | 4780 | 17:35:16 | 7.4 | 12.9 | 8770 | 4280 | 16:58:26 | 6.8 | 13.8 | 8770 | 4230 | VL | Io-C | 0.34 |
| 32 | 2008-11-11 | - | - | - | - | - | 03:12:01 | 2.6 | 9 | 9750 | 4680 | 02:26:18 | 4.8 | 23.7 | 9930 | 4150 | VL | Undefined | 0.34 |
| 33* | 2008-11-15 | 15:11:06 | 8.2 | 22.6 | 7820 | 12780 | 15:45:24 | 7.2 | 19.9 | 8030 | 12640 | 14:58:04 | 6.5 | 20.6 | 7940 | 12780 | VE | Io-D | 0.36 |
| 34 | 2008-11-18 | 04:23:24 | 8.8 | 14.9 | 7820 | 3840 | 04:58:00 | 7.1 | 14.6 | 7940 | 3920 | 04:01:21 | 8.5 | 16 | 7940 | 4230 | VL | Undefined | 0.30 |
| 35* | 2008-11-22 | 16:30:06 | 9.8 | 17.1 | 4980 | 11390 | 16:53:48 | 10.5 | 14.3 | 4780 | 10130 | 16:05:21 | 9.8 | 21.6 | 4940 | 11440 | VE | Io-D | 0.38 |
| 36* | 2008-12-17 | 11:57:28 | 7.8 | 17.4 | 4930 | 11270 | 12:22:39 | 8.5 | 16.3 | 4940 | 11440 | 11:26:43 | 5.9 | 28.2 | 4990 | 12500 | VE | Io-D | 0.40 |
| 37* | 2008-12-24 | 13:11:30 | 4.2 | 18.5 | 4890 | 10510 | 14:02:00 | 7.4 | 30.2 | 4920 | 9560 | 12:47:34 | 8 | 18.7 | 4950 | 12260 | VE | Io-D | 0.30 |
| 38 | 2008-12-25 | 09:14:28 | 6.9 | 18.5 | 7820 | 3760 | 09:47:21 | 7.2 | 15 | 7940 | 3540 | - | - | - | - | - | VL | Io-C | 0.34 |
| 39 | 2009-01-01 | 10:49:47 | 6.5 | 13.9 | 5890 | 12520 | - | - | - | - | - | 10:08:03 | 7.1 | 16.5 | 5830 | 12220 | VL | Undefined | 0.28 |
| 40 | 2009-02-02 | 07:56:02 | 7.8 | 15.2 | 5890 | 11390 | 08:26:32 | 7.4 | 33.6 | 5960 | 12780 | 07:13:42 | 7.4 | 12 | 5830 | 11960 | VL | Undefined | 0.31 |
| 41 | 2009-02-07 | 15:49:24 | 6.5 | 20.8 | 6020 | 3180 | 16:19:04 | 7.6 | 23.2 | 6030 | 2630 | - | - | - | - | - | VL | Undefined | 0.30 |
| 42* | 2009-03-03 | 15:14:05 | 10.6 | 25 | 4980 | 10590 | 15:29:52 | 11.3 | 10.6 | 4880 | 10360 | 14:40:07 | 8.6 | 26.1 | 4940 | 10820 | VE | Io-D | 0.36 |
| 43 | 2009-03-10 | 17:20:46 | 5.9 | 20.6 | 3870 | 10020 | 17:44:45 | 10.1 | 18.5 | 3960 | 12800 | 16:01:30 | 7.8 | 9.9 | 3980 | 11150 | VE | Io-D | 0.17 |
| 44 | 2009-03-16 | 20:26:50 | 5.4 | 18.3 | 7820 | 3640 | 20:41:32 | 7.2 | 18 | 8030 | 3790 | - | - | - | - | - | VL | Io-C | 0.29 |
| 45 | 2009-03-23 | 22:10:24 | 6.8 | 23.4 | 6020 | 2270 | 22:20:48 | 7.5 | 16.6 | 5960 | 2690 | 21:35:17 | 7.5 | 13.3 | 6030 | 3920 | VL | Undefined | 0.31 |
| 46 | 2009-03-28 | 00:34:46 | 5.7 | 28.3 | 5980 | 9440 | 00:41:48 | 4.2 | 12.6 | 6020 | 8330 | - | - | - | - | - | VE | Undefined | 0.30 |
| 47* | 2009-03-28 | 10:32:21 | 6.6 | 15.3 | 4910 | 11510 | 10:30:54 | 6.7 | 20.2 | 5150 | 12140 | 10:04:32 | 6 | 22.7 | 5020 | 10470 | VE | Io-D | 0.43 |
| 48* | 2009-05-19 | 13:28:45 | 6.9 | 14.7 | 6080 | 11760 | 12:49:03 | 5.7 | 15.6 | 6030 | 12220 | 13:21:14 | 6.6 | 15.3 | 5890 | 12500 | VL | Io-C | 0.28 |
| 49* | 2009-05-31 | 03:30:08 | 5.2 | 18.8 | 4910 | 11280 | 02:50:35 | 8.4 | 22 | 4990 | 10820 | 03:31:51 | 4.5 | 19.3 | 5020 | 10810 | VE | Io-D | 0.37 |
| 50* | 2009-05-31 | 22:47:25 | 4.8 | 8.9 | 8890 | 12340 | 21:59:54 | 4.4 | 10.7 | 8780 | 13910 | 22:46:56 | 4.4 | 8.2 | 9040 | 13760 | VL | Io-C | 0.31 |
| 51 | 2009-06-07 | 04:34:57 | 8.6 | 7.2 | 4910 | 10220 | 04:07:16 | 4.1 | 6.7 | 4000 | 10490 | 04:41:09 | 4.1 | 8.2 | 3780 | 9930 | VE | Undefined | 0.64 |

| | | | | | | | | | | | | | | | | | | |
|---|---|---|---|---|---|---|---|---|---|---|---|---|---|---|---|---|---|---|
| 52* | 2009-06-12 | 12:42:38 | 4.4 | 7 | 3950 | 10220 | 12:01:40 | 7.5 | 25 | 3960 | 12400 | 12:51:03 | 3.7 | 9.6 | 4020 | 10920 | VE | Io-D | 0.44 |
| 53 | 2009-07-22 | 06:08:50 | 8.7 | 20.4 | 7820 | 3250 | 04:53:11 | 11.3 | 27.8 | 8120 | 3540 | 06:54:02 | 11.8 | 24.8 | 7940 | 2570 | VL | Undefined | 0.28 |
| 54 | 2009-08-02 | - | - | - | - | - | 18:47:41 | 6.3 | 10 | 4000 | 8950 | 20:12:51 | 3.8 | 4.7 | 3980 | 9030 | VE | Undefined | 0.41 |
| 55* | 2009-08-02 | 19:41:12 | 7.6 | 9.6 | 4030 | 10630 | 19:00:42 | 7.6 | 21.7 | 3960 | 13200 | 20:36:21 | 10.3 | 36.2 | 3900 | 9630 | VE | Io-D | 0.37 |
| 56* | 2009-08-13 | 02:48:00 | 6 | 19 | 6000 | 11000 | 02:28:00 | 7 | 21 | 6000 | 11000 | 03:13:00 | 10 | 26 | 6000 | 11000 | VL | Non-Io-C | 0.81 |
| 57 | 2009-08-23 | 02:04:14 | 7.7 | 26.7 | 6820 | 2630 | 01:14:36 | 7 | 21.2 | 7050 | 3630 | 03:16:05 | 7.4 | 14 | 6730 | 3170 | VL | Io-C | 0.28 |
| 58* | 2009-08-27 | 14:21:53 | 12.6 | 22.3 | 4980 | 12000 | - | - | - | - | - | 15:17:35 | 7 | 26 | 4830 | 13800 | VE | Io-D | 0.38 |
| 59* | 2009-09-03 | 15:32:14 | 6.8 | 21.7 | 5010 | 10950 | 15:07:36 | 5.9 | 32.7 | 4830 | 12400 | 17:04:55 | 15.8 | 22.2 | 4970 | 8940 | VE | Io-D | 0.27 |
| 60* | 2009-09-11 | 12:48:17 | 6.1 | 12.2 | 7730 | 3760 | 12:27:22 | 5.2 | 12.5 | 8120 | 4180 | - | - | - | - | - | VL | Io-C | 0.30 |
| 61* | 2009-09-23 | 21:46:17 | 11.6 | 17 | 6890 | 12780 | 21:36:36 | 9.3 | 15.9 | 7030 | 13800 | 23:11:01 | 10.8 | 35.6 | 6730 | 13800 | VL | Io-C | 0.26 |
| 62* | 2009-09-28 | 10:32:49 | 7.8 | 20.8 | 6010 | 12270 | 10:37:04 | 7.2 | 23.2 | 5920 | 13690 | 11:23:32 | 9 | 27.6 | 6030 | 13090 | VE | Io-D | 0.43 |
| 63* | 2010-03-12 | 02:13:00 | 7.3 | 22.7 | 6020 | 11760 | 02:43:25 | 9.5 | 7 | 5830 | 10240 | 01:17:38 | 6.2 | 32.6 | 6090 | 13650 | VE | Io-D | 0.35 |
| 64 | 2010-03-19 | 03:20:15 | 8.1 | 42.7 | 4980 | 10370 | - | - | - | - | - | 02:29:06 | 9.2 | 32.7 | 4940 | 12090 | VE | Io-D | 0.38 |
| 65 | 2010-04-25 | 08:24:26 | 8.3 | 36.7 | 4360 | 11070 | 08:34:05 | 8.7 | 31.8 | 4410 | 12820 | 07:33:46 | 5.9 | 39.9 | 4390 | 12260 | VE | Io-D | 0.32 |
| 66 | 2010-06-28 | 01:52:56 | 8.8 | 37.1 | 4930 | 11390 | 00:26:58 | 6.3 | 61.5 | 4880 | 13650 | 01:08:15 | 7.6 | 50.6 | 4830 | 12360 | VE | Io-D | 0.25 |
| 67* | 2010-07-05 | 03:06:38 | 3.9 | 3.1 | 4940 | 13060 | 02:13:05 | 4.4 | 14.7 | 4760 | 13100 | - | - | - | - | - | VE | Io-D | 0.47 |
| 68 | 2010-07-10 | 11:19:36 | 8.2 | 26 | 4910 | 11970 | 09:49:53 | 6.7 | 57.3 | 4830 | 13480 | 10:58:52 | 10.3 | 46.3 | 4990 | 13060 | VE | Io-D | 0.28 |
| 69 | 2010-07-29 | 22:14:34 | 9.4 | 30.2 | 4980 | 11390 | 20:17:54 | 14.6 | 32.1 | 5050 | 14750 | 22:21:11 | 9.4 | 29.8 | 4830 | 11570 | VE | Io-D | 0.24 |
| 70* | 2010-08-06 | 02:56:22 | 6.8 | 19.9 | 3910 | 9720 | 02:26:35 | 6.9 | 20.4 | 3880 | 9540 | 03:10:23 | 7.4 | 18.9 | 4020 | 9930 | VL | Non-Io-A | 0.95 |
| 71* | 2010-08-06 | 03:10:43 | 6.8 | 20 | 3870 | 10020 | 02:42:11 | 8.4 | 25.4 | 3880 | 10600 | 03:26:30 | 6.2 | 21 | 3980 | 10580 | VL | Non-Io-A | 0.99 |
| 72 | 2010-08-23 | - | - | - | - | - | 15:15:47 | 7.7 | 52.3 | 4830 | 14210 | 17:10:50 | 8.3 | 16.6 | 5020 | 10810 | VE | Undefined | 0.33 |
| 73* | 2010-08-30 | 18:38:28 | 4.4 | 23.8 | 4490 | 12710 | 16:35:02 | 6.2 | 35.7 | 4490 | 12800 | 19:15:38 | 8.2 | 15.6 | 4420 | 13190 | VE | Io-D | 0.44 |
| 74* | 2010-09-12 | 03:31:22 | 5.5 | 25.7 | 4910 | 11860 | 02:26:00 | 5.1 | 35 | 4830 | 13200 | - | - | - | - | - | VE | Io-D | 0.30 |
| 75* | 2010-10-26 | 08:22:34 | 11.2 | 26.8 | 3910 | 10950 | 08:28:59 | 7.8 | 26.1 | 4990 | 11280 | 10:24:54 | 9.5 | 36.6 | 4860 | 12000 | VE | Io-D | 0.27 |
| 76* | 2010-11-27 | 04:38:31 | 3.6 | 25.2 | 4910 | 10630 | 05:15:44 | 5.9 | 29.3 | 4830 | 11050 | 06:09:03 | 5.7 | 21.1 | 4810 | 10470 | VE | Io-D | 0.30 |
| 77* | 2010-12-09 | 14:20:32 | 4.6 | 19.2 | 5980 | 12580 | 14:53:04 | 6.2 | 30.8 | 5900 | 11400 | 15:14:26 | 6.7 | 20.2 | 6060 | 12000 | VE | Io-D | 0.42 |
| 78 | 2011-02-18 | 09:33:56 | 5.6 | 31.8 | 3910 | 10950 | - | - | - | - | - | 09:08:04 | 6.1 | 32.2 | 3900 | 10920 | VE | Io-D | 0.33 |
| 79 | 2011-06-24 | 03:32:45 | 8.3 | 25.4 | 3960 | 10370 | 02:40:14 | 7.2 | 37.2 | 3870 | 11820 | 01:53:31 | 8.1 | 24.4 | 3870 | 10240 | VE | Io-D | 0.23 |
| 80* | 2011-08-07 | 09:32:27 | 4.9 | 23.2 | 4910 | 12340 | 07:45:30 | 4.5 | 27.7 | 4940 | 12930 | 08:49:12 | 6.8 | 37.1 | 5020 | 12390 | VE | Io-D | 0.29 |
| 81 | 2011-11-16 | 05:26:26 | 5.9 | 24.3 | 4760 | 10630 | - | - | - | - | - | 06:48:20 | 7.7 | 44.7 | 5020 | 13620 | VE | Io-D | 0.37 |
| 82* | 2012-07-26 | 09:16:11 | 9 | 18 | 4860 | 10630 | 08:12:32 | 6.8 | 28 | 4990 | 11640 | - | - | - | - | - | VE | Io-D | 0.34 |
| 83* | 2012-10-29 | 22:35:57 | 7.6 | 23.1 | 6810 | 13090 | 21:06:35 | 6.2 | 24.4 | 6900 | 12020 | 22:18:45 | 8.1 | 38.6 | 7100 | 12650 | VE | Io-D | 0.30 |
| 84 | 2013-06-15 | - | - | - | - | - | 06:28:42 | 7.6 | 20.8 | 4940 | 11400 | 04:28:07 | 5.6 | 29.5 | 4970 | 12000 | VE | Undefined | 0.23 |
| 85 | 2013-07-17 | 02:44:44 | 14.3 | 28.6 | 5040 | 9740 | 02:25:21 | 14.8 | 44.8 | 4940 | 10820 | 00:32:47 | 12.9 | 43.4 | 4830 | 13210 | VE | Io-D | 0.20 |
| 86 | 2013-07-29 | 11:39:09 | 9 | 55.6 | 5040 | 12130 | 10:55:10 | 8.5 | 66 | 4940 | 12500 | 10:11:11 | 10.8 | 37.2 | 4940 | 13800 | VE | Io-D | 0.35 |
| 87* | 2013-07-30 | 06:57:18 | 9.4 | 18.4 | 6890 | 12920 | 06:09:38 | 12.2 | 20.2 | 6880 | 13350 | 04:31:38 | 10.8 | 21.6 | 6960 | 11960 | VL | Io-C | 0.26 |
| 88 | 2013-08-10 | 21:40:57 | 13.3 | 40.9 | 5040 | 12390 | 20:27:50 | 9.2 | 60.7 | 4880 | 13800 | 19:54:44 | 12.4 | 29 | 4830 | 13960 | VE | Io-D | 0.25 |
| 89 | 2013-08-17 | - | - | - | - | - | 22:07:46 | 7.8 | 47.7 | 5830 | 14270 | 21:39:20 | 9.8 | 35.2 | 5890 | 13060 | VE | Undefined | 0.33 |
| 90* | 2013-08-30 | 08:36:58 | 8 | 10 | 4980 | 11760 | 07:21:13 | 12 | 36.7 | 4880 | 13960 | 06:57:18 | 7.2 | 18 | 4830 | 11440 | VE | Io-D | 0.33 |
| 91 | 2013-09-11 | 18:02:59 | 13.2 | 26.5 | 6820 | 12920 | 16:28:28 | 10.3 | 39.3 | 6880 | 13060 | 16:34:01 | 10.3 | 27.4 | 6960 | 12640 | VE | Undefined | 0.35 |
| 92 | 2013-09-18 | 17:04:37 | 4.2 | 14.1 | 4910 | 10950 | 16:24:24 | 5.3 | 12.8 | 4830 | 11050 | 16:34:43 | 4.2 | 14.7 | 5020 | 11510 | VL | Non-Io-C | 0.97 |
| 93* | 2013-09-18 | 17:21:48 | 5.8 | 16 | 3990 | 8300 | 16:42:15 | 3.9 | 15.9 | 3880 | 9150 | 16:50:40 | 4.5 | 17.4 | 3900 | 10920 | VL | Non-Io-C | 0.99 |
| 94 | 2013-09-18 | 17:35:06 | 5 | 18.6 | 3910 | 8810 | 16:55:55 | 4.2 | 15.9 | 3880 | 7740 | 17:04:58 | 5.2 | 20.8 | 3860 | 9230 | VL | Non-Io-C | 1.00 |
| 95 | 2013-09-24 | 21:13:27 | 5.3 | 18.8 | 3980 | 9580 | 20:32:49 | 9.1 | 31.9 | 4000 | 11880 | 20:46:02 | 6.2 | 31.2 | 3900 | 11240 | VE | Non-Io-D | 0.96 |
| 96* | 2013-09-24 | 23:18:02 | 6.7 | 22.4 | 2930 | 7440 | 22:33:45 | 6.4 | 17.7 | 2920 | 7820 | 22:46:10 | 6.1 | 14.8 | 3030 | 6880 | VL | Non-Io-C | 0.87 |
| 97 | 2013-10-20 | - | - | - | - | - | 14:25:29 | 6.5 | 35.2 | 5840 | 13480 | 15:07:21 | 7.6 | 24.2 | 5870 | 11880 | VE | Undefined | 0.26 |
| 98 | 2013-11-14 | 11:28:14 | 6.1 | 10.9 | 6810 | 12710 | 09:13:55 | 7.9 | 39.6 | 6900 | 13340 | 10:06:42 | 9.7 | 40.3 | 6880 | 12920 | VE | Io-D | 0.23 |
| 99 | 2013-11-14 | - | - | - | - | - | 09:28:50 | 3.2 | 7.2 | 6900 | 2070 | 10:34:20 | 4.7 | 6 | 6880 | 3130 | VE | Undefined | 0.22 |
| 100 | 2013-12-16 | - | - | - | - | - | 06:16:48 | 8.7 | 50.7 | 5960 | 12660 | 07:24:02 | 7.5 | 14.2 | 5870 | 9230 | VE | Undefined | 0.23 |
| 101* | 2014-01-02 | 23:54:12 | 5.7 | 28.6 | 5870 | 11280 | 23:56:06 | 5.5 | 28 | 6080 | 12660 | 00:29:20 | 7.1 | 32.2 | 6000 | 11750 | VE | Io-D | 0.47 |
| 102* | 2014-02-03 | 20:04:26 | 6.7 | 22 | 4910 | 12960 | - | - | - | - | - | 21:04:57 | 9.6 | 26 | 5020 | 11630 | VE | Io-D | 0.48 |
| 103 | 2014-02-16 | - | - | - | - | - | 06:21:44 | 9.6 | 32.6 | 4940 | 12530 | 06:53:02 | 6.5 | 31 | 4860 | 12130 | VE | Undefined | 0.36 |
| 104* | 2014-03-14 | 17:22:00 | 10 | 15 | 5000 | 16000 | 19:24:00 | 10 | 15 | 5000 | 16000 | 19:44:00 | 10 | 15 | 5000 | 16000 | VE | Io-D | 0.25 |
| 105* | 2014-03-27 | 02:55:12 | 5.5 | 27.7 | 3910 | 11280 | 04:12:22 | 8.3 | 12.6 | 4000 | 9150 | 05:19:59 | 9 | 44.2 | 4020 | 12000 | VE | Io-D | 0.26 |
| 106* | 2014-05-10 | 06:07:54 | 4.4 | 27 | 4910 | 11280 | 06:40:37 | 3.8 | 27 | 4990 | 11770 | 06:34:32 | 3.8 | 33.4 | 5020 | 12780 | VL | Non-Io-A | 0.90 |
| 107 | 2014-05-10 | 09:15:06 | 6.6 | 22.3 | 3990 | 11620 | 11:00:45 | 8.4 | 26.2 | 4090 | 8330 | 10:48:30 | 8.6 | 39.8 | 3900 | 9630 | VE | Io-D | 0.30 |
| 108* | 2014-07-06 | 01:26:36 | 4.8 | 18.8 | 5980 | 12710 | 01:54:59 | 4.5 | 22.1 | 5960 | 12930 | 01:23:27 | 6 | 28.2 | 6060 | 13760 | VE | Io-D | 0.39 |
| 109* | 2014-08-06 | 21:58:13 | 6.6 | 27.2 | 5530 | 12340 | 21:34:16 | 7.7 | 35.6 | 5900 | 13620 | 21:01:35 | 6.1 | 29.5 | 5870 | 12920 | VE | Io-D | 0.36 |
| 110 | 2014-08-19 | 07:37:21 | 5 | 28.2 | 6040 | 11280 | 06:49:18 | 7.7 | 44.3 | 5900 | 12800 | 06:33:16 | 7 | 34.8 | 5890 | 13800 | VE | Io-D | 0.39 |
| 111 | 2014-09-15 | 14:00:34 | 5.3 | 20.6 | 3910 | 11170 | - | - | - | - | - | 13:15:57 | 5.8 | 19.9 | 3900 | 10810 | VL | Non-Io-C | 0.83 |
| NDA events | | | | | | | | | | | | | | | | | | | |
| 112* | 2008-03-02 | 07:54:00 | 8 | 20 | 21000 | 31400 | - | - | - | - | - | - | - | - | - | - | VL | Io-B | - |
| 113* | 2008-08-20 | 22:24:00 | 4 | 9 | 24000 | 31600 | - | - | - | - | - | - | - | - | - | - | VL | Io-B | - |
| 114* | 2008-10-23 | 15:00:00 | 8 | 14 | 22000 | 30000 | - | - | - | - | - | - | - | - | - | - | VL | Io-B | - |
| 115* | 2008-10-30 | 16:15:00 | 7 | 10 | 27000 | 34000 | - | - | - | - | - | - | - | - | - | - | VL | Io-B | - |
| 116* | 2009-08-24 | 00:37:00 | 24 | 30 | 22700 | 33000 | - | - | - | - | - | - | - | - | - | - | VE | Io-A | - |
| 117* | 2008-06-10 | 02:41:00 | 4 | -5 | 21500 | 30000 | - | - | - | - | - | - | - | - | - | - | VE | Io-A | - |
| 118* | 2009-05-30 | 03:46:00 | 5 | -16 | 18000 | 28000 | - | - | - | - | - | - | - | - | - | - | VE | Io-A | - |
| 119* | 2010-06-02 | 05:47:00 | 5 | -7 | 20000 | 30000 | - | - | - | - | - | - | - | - | - | - | VE | Io-A | - |
| 120* | 2010-07-11 | 03:02:00 | 4 | -12 | 18700 | 31000 | - | - | - | - | - | - | - | - | - | - | VE | Io-A | - |
| 121* | 2010-08-03 | 06:38:00 | 6 | -15 | 20000 | 31000 | - | - | - | - | - | - | - | - | - | - | VL | Io-C | - |

Column (1) is the number of events, which marked with an asterisk indicates that the event is used for the method. The events with number 112 to 121 are observed from Nançay Decameter Array (NDA). Column (2) is the observed time in year-month-day format. Column (3) shows the first observed moment at the apex of DAM arc from the corresponding observer. Column (4) is the observed period at the vertex frequency. Column (5) indicates the time span from the moment corresponding to the vertex frequency to the moment at the end (high or low frequency), in terms of the frontier of the arc, where column (6) represents the vertex frequency and column (7) means the frequency at the end, respectively. Columns (8)-(17) have the same meanings of Columns (3)-(7). Column (18) indicates that the DAM is the vertex-early/late (or VE/VL) arc. Column (19) means the types for DAMs. Column (20) is the apparent rotation speed calculated by Eq.1 based on multi-view observations.

**Table 2.** Statistics of emission parameters for model

| No. | Longitude of Footprint [°] | | Latitude of Footprint [°] | | M-shell value [$R_J$] | | Emission Angle [°] | | | Energy of Electron [keV] | | |
|---|---|---|---|---|---|---|---|---|---|---|---|---|
| | 10% | 90% | 10% | 90% | 10% | 90% | Avg | Med | Std | Avg | Med | Std |
| 1 | 328.28 | 294.27 | -75.26 | -49.11 | 2.52 | 11.75 | 81.39 | 81.61 | 3.45 | 1.56 | 1.25 | 1.22 |
| 3 | 74.41 | 47.08 | -69.06 | -44.74 | 3.54 | 18.84 | 64.91 | 64.38 | 5.81 | 12.49 | 12.94 | 4.89 |
| 4 | 88.02 | 71.48 | -68.19 | -55.39 | 4.27 | 12.97 | 63.48 | 63.23 | 4.88 | 13.30 | 13.71 | 3.72 |
| 6 | 276.81 | 238.90 | -77.23 | -54.03 | 2.38 | 6.51 | 69.11 | 68.91 | 6.98 | 8.85 | 8.03 | 5.14 |
| 9 | 66.93 | 46.74 | -67.12 | -46.54 | 3.65 | 19.04 | 62.84 | 62.85 | 4.81 | 13.16 | 13.30 | 3.06 |
| 13 | 85.63 | 62.91 | -68.09 | -49.88 | 4.14 | 14.49 | 63.48 | 63.19 | 4.87 | 13.20 | 13.37 | 3.67 |
| 14 | 315.38 | 258.70 | -76.26 | -52.39 | 2.46 | 14.21 | 83.69 | 84.15 | 2.63 | 0.90 | 0.63 | 0.83 |
| 15 | 299.48 | 265.60 | -76.61 | -52.16 | 2.42 | 6.76 | 65.40 | 65.13 | 5.72 | 9.62 | 9.26 | 3.80 |
| 17 | 68.38 | 40.97 | -67.55 | -48.97 | 4.79 | 17.89 | 62.52 | 62.54 | 4.57 | 12.98 | 12.76 | 3.06 |
| 20 | 75.44 | 49.61 | -67.55 | -56.87 | 3.75 | 17.89 | 65.94 | 65.20 | 5.71 | 10.55 | 10.51 | 4.51 |
| 21 | 104.64 | 66.53 | -67.14 | -48.49 | 3.94 | 12.70 | 62.54 | 62.40 | 4.58 | 13.09 | 13.33 | 3.19 |
| 23 | 75.57 | 44.11 | -67.27 | -45.46 | 3.74 | 14.35 | 62.06 | 61.85 | 4.42 | 12.40 | 12.39 | 2.71 |
| 24 | 107.69 | 82.54 | -67.91 | -58.93 | 4.59 | 9.29 | 62.57 | 62.51 | 4.41 | 12.04 | 12.51 | 2.36 |
| 25 | 77.40 | 60.25 | -64.18 | -45.46 | 3.73 | 10.87 | 61.77 | 61.54 | 4.35 | 10.35 | 10.40 | 1.72 |
| 26 | 295.95 | 249.60 | -77.10 | -53.42 | 2.52 | 6.78 | 69.66 | 69.09 | 7.53 | 7.35 | 7.11 | 4.24 |
| 27 | 62.33 | 44.11 | -66.40 | -46.42 | 3.47 | 15.34 | 61.96 | 61.91 | 4.19 | 12.55 | 12.26 | 2.61 |
| 28 | 136.75 | 115.60 | -72.93 | -57.28 | 2.93 | 6.32 | 64.01 | 63.59 | 5.51 | 10.41 | 11.09 | 3.14 |
| 31 | 328.28 | 304.07 | -75.90 | -63.46 | 5.64 | 52.54 | 84.10 | 84.87 | 3.00 | 0.89 | 0.51 | 1.11 |
| 33 | 64.56 | 48.01 | -62.00 | -58.05 | 6.44 | 9.39 | 59.77 | 59.50 | 3.34 | 12.65 | 12.65 | 1.16 |
| 35 | 73.78 | 54.48 | -67.01 | -58.30 | 7.22 | 18.59 | 61.50 | 61.51 | 3.80 | 14.07 | 14.19 | 2.10 |
| 36 | 73.30 | 46.83 | -67.19 | -54.33 | 5.79 | 18.59 | 62.20 | 62.29 | 4.23 | 13.27 | 13.46 | 2.54 |
| 37 | 77.57 | 65.06 | -67.14 | -53.61 | 5.73 | 14.00 | 62.41 | 62.26 | 4.33 | 12.51 | 12.78 | 2.31 |
| 42 | 90.00 | 76.18 | -65.60 | -56.49 | 5.59 | 9.91 | 61.39 | 61.37 | 3.94 | 14.06 | 14.41 | 2.05 |
| 47 | 82.00 | 52.34 | -67.14 | -46.36 | 4.07 | 14.73 | 62.76 | 62.87 | 4.42 | 12.34 | 12.57 | 2.65 |
| 48 | 337.40 | 299.77 | -74.47 | -48.41 | 2.53 | 9.88 | 71.98 | 72.09 | 5.65 | 5.67 | 5.04 | 3.40 |
| 49 | 85.20 | 65.88 | -66.07 | -56.69 | 5.67 | 11.99 | 61.79 | 61.99 | 3.87 | 13.41 | 13.68 | 2.07 |
| 50 | 314.42 | 273.68 | -73.28 | -51.19 | 2.46 | 6.99 | 75.95 | 76.31 | 6.43 | 3.22 | 2.53 | 2.65 |
| 52 | 81.78 | 50.11 | -67.08 | -48.05 | 3.75 | 15.19 | 68.63 | 68.63 | 5.97 | 8.47 | 8.26 | 3.96 |
| 55 | 88.09 | 65.03 | -68.17 | -46.47 | 3.92 | 13.13 | 64.38 | 64.50 | 5.04 | 11.31 | 11.41 | 2.95 |
| 56 | 315.23 | 286.95 | -73.28 | -51.20 | 2.50 | 7.58 | 62.98 | 62.78 | 4.66 | 12.44 | 12.24 | 3.27 |
| 58 | 80.09 | 51.82 | -65.06 | -57.14 | 6.05 | 12.00 | 62.36 | 62.24 | 4.63 | 12.18 | 11.97 | 2.44 |
| 59 | 72.97 | 57.87 | -64.38 | -60.25 | 7.65 | 11.99 | 60.82 | 60.84 | 3.53 | 14.26 | 14.35 | 1.41 |
| 60 | 335.27 | 311.24 | -75.54 | -65.17 | 6.45 | 51.71 | 84.82 | 85.37 | 2.31 | 0.67 | 0.42 | 0.67 |
| 61 | 368.77 | 293.81 | -71.69 | -50.00 | 2.57 | 9.83 | 74.15 | 75.19 | 8.19 | 4.65 | 3.20 | 4.26 |
| 62 | 76.00 | 48.76 | -65.74 | -52.45 | 5.09 | 14.49 | 62.50 | 62.39 | 4.61 | 11.58 | 11.34 | 2.61 |
| 63 | 90.00 | 51.74 | -64.04 | -58.88 | 4.09 | 10.51 | 61.77 | 61.51 | 4.28 | 13.12 | 13.24 | 2.81 |
| 67 | 143.10 | 103.82 | -71.47 | -49.54 | 2.76 | 6.68 | 69.71 | 69.52 | 7.12 | 6.93 | 6.94 | 3.60 |
| 70 | 162.57 | 150.12 | 47.04 | 54.95 | 5.70 | 14.13 | 65.55 | 64.78 | 6.90 | 14.89 | 14.72 | 7.34 |
| 71 | 166.32 | 156.69 | 47.77 | 54.39 | 6.62 | 16.58 | 64.44 | 63.51 | 6.06 | 15.82 | 16.21 | 6.67 |
| 73 | 138.12 | 117.14 | -67.60 | -60.38 | 3.27 | 4.90 | 63.63 | 63.86 | 4.92 | 11.52 | 11.89 | 2.28 |
| 74 | 92.17 | 67.52 | -65.13 | -60.25 | 6.10 | 8.62 | 61.76 | 61.93 | 4.17 | 13.03 | 13.36 | 1.31 |
| 75 | 62.42 | 52.77 | -64.83 | -61.09 | 8.59 | 14.35 | 60.83 | 60.90 | 3.03 | 15.08 | 15.02 | 1.45 |
| 76 | 94.38 | 58.15 | -65.18 | -59.38 | 6.55 | 14.00 | 61.44 | 61.66 | 3.47 | 14.12 | 14.12 | 1.57 |
| 77 | 82.13 | 56.79 | -64.00 | -47.62 | 3.69 | 11.56 | 63.04 | 63.09 | 4.70 | 11.04 | 11.17 | 2.69 |
| 80 | 111.07 | 71.10 | -65.37 | -56.14 | 4.71 | 10.10 | 62.46 | 62.76 | 4.46 | 11.87 | 12.19 | 1.83 |
| 82 | 82.39 | 63.32 | -64.54 | -58.53 | 6.06 | 9.31 | 61.31 | 61.36 | 3.89 | 13.91 | 14.17 | 1.81 |
| 83 | 104.19 | 64.33 | -62.25 | -59.22 | 4.97 | 8.63 | 60.13 | 60.18 | 3.69 | 12.62 | 12.59 | 1.11 |
| 87 | 326.53 | 247.69 | -75.23 | -50.27 | 2.56 | 7.61 | 72.37 | 73.01 | 7.29 | 5.72 | 4.36 | 4.60 |
| 90 | 119.68 | 73.81 | -67.07 | -52.22 | 3.30 | 8.73 | 63.21 | 63.28 | 4.77 | 11.85 | 12.26 | 2.63 |
| 93 | 323.89 | 295.04 | -73.92 | -50.27 | 2.56 | 8.30 | 65.53 | 65.37 | 5.48 | 11.68 | 11.17 | 4.60 |
| 96 | 358.46 | 330.63 | -73.15 | -46.79 | 2.78 | 14.41 | 64.73 | 64.14 | 5.86 | 13.85 | 14.01 | 5.39 |
| 101 | 58.72 | 48.88 | -64.02 | -60.69 | 8.30 | 10.96 | 60.50 | 60.47 | 3.36 | 14.34 | 14.12 | 2.19 |

| | | | | | | | | | | | |
|---|---|---|---|---|---|---|---|---|---|---|---|
| 102 | 73.82 | 49.59 | -66.14 | -44.72 | 3.74 | 12.27 | 63.21 | 63.39 | 4.94 | 11.38 | 11.37 | 2.59 |
| 104 | 138.98 | 75.05 | -69.91 | -53.63 | 3.65 | 12.97 | 69.97 | 70.39 | 5.19 | 5.75 | 5.62 | 1.83 |
| 105 | 84.27 | 58.48 | -65.51 | -57.43 | 5.85 | 12.50 | 61.91 | 62.02 | 3.86 | 14.57 | 14.45 | 2.33 |
| 106 | 188.04 | 184.67 | 55.49 | 56.43 | 8.95 | 17.01 | 59.69 | 59.50 | 2.86 | 20.41 | 20.82 | 2.91 |
| 108 | 86.90 | 67.42 | -65.36 | -48.31 | 3.99 | 11.06 | 63.10 | 63.25 | 4.58 | 10.60 | 10.73 | 2.00 |
| 109 | 73.99 | 51.03 | -63.74 | -60.22 | 8.37 | 12.00 | 60.42 | 60.23 | 3.51 | 14.05 | 13.69 | 1.97 |
| 112 | 190.25 | 170.02 | 33.11 | 55.55 | 1.60 | 7.43 | 65.56 | 65.07 | 6.60 | 6.62 | 6.56 | 3.44 |
| 113 | 191.78 | 166.55 | 33.11 | 57.34 | 1.68 | 7.00 | 70.63 | 71.05 | 7.76 | 4.48 | 3.61 | 3.45 |
| 114 | 188.60 | 164.84 | 30.80 | 56.37 | 1.55 | 7.00 | 66.78 | 66.18 | 7.10 | 6.30 | 6.33 | 3.66 |
| 115 | 210.30 | 174.40 | 39.24 | 60.94 | 1.91 | 6.42 | 71.20 | 71.57 | 8.15 | 3.41 | 2.54 | 2.89 |
| 116 | 185.10 | 174.40 | 49.49 | 53.83 | 5.64 | 7.51 | 62.35 | 62.15 | 4.34 | 8.76 | 8.96 | 1.50 |
| 117 | 171.52 | 151.39 | 23.49 | 52.47 | 1.35 | 5.70 | 75.25 | 75.97 | 6.94 | 3.13 | 2.23 | 2.75 |
| 118 | 182.03 | 161.27 | 22.84 | 53.07 | 1.33 | 8.21 | 66.47 | 65.61 | 7.13 | 7.41 | 7.42 | 4.04 |
| 119 | 172.07 | 156.91 | 23.60 | 52.47 | 1.38 | 6.10 | 73.90 | 74.39 | 6.87 | 3.69 | 2.87 | 2.90 |
| 120 | 177.43 | 163.43 | 25.34 | 53.83 | 1.45 | 8.03 | 71.26 | 71.20 | 6.88 | 4.72 | 4.16 | 3.04 |
| 121 | 179.08 | 161.40 | 24.47 | 53.06 | 1.41 | 8.18 | 67.86 | 67.25 | 7.26 | 6.17 | 6.21 | 3.41 |

The longitude and latitude of FPs are expressed as the range between 10 and 90 percent. The M-shell value represents the distance at the top of field lines (or at the magnetic equator) from the Jupiter.